\RequirePackage{fix-cm}
\documentclass[twocolumn,epjc3]{svjour3}  
\smartqed  
\RequirePackage{graphicx}
\usepackage{amsmath}
\usepackage{mathtools}
\usepackage{amssymb}
\usepackage{slashed}
\usepackage{braket}
\usepackage{revsymb}
\usepackage{subcaption}
\usepackage{cuted}
\usepackage{hyperref}

\journalname{Eur. Phys. J. C}
\begin{document}

\title{Neutron tagging of heavy lepton pair production  at RHIC and LHC energies
}


\author{Atacan Fatih Candar\thanksref{e1,addr1}
        \and
        Mehmet Cem Güçlü\thanksref{e2,addr1} 
}

\thankstext{e1}{e-mail: candara15@itu.edu.tr}
\thankstext{e2}{e-mail: guclu@itu.edu.tr}

\institute{Department of Physics, İstanbul Technical University, İstanbul, Türkiye \label{addr1}        
}

\date{Received: date / Accepted: date}

\maketitle

\begin{abstract}
We calculated cross sections for $\gamma\gamma \rightarrow \ell^+\ell^-$ ($\ell=e,\mu,\tau$)  production in ultraperipheral gold and lead collisions and compare our results with recent measurements from the STAR, ALICE, ATLAS, and CMS experiments at center of mass energies $\sqrt{s_{NN}}=200$ GeV, $\sqrt{s_{NN}}=2.76 $ TeV, and $\sqrt{s_{NN}}=5.02$ TeV. The calculations are performed within lowest order quantum electrodynamics (LO QED) using a Monte Carlo technique for calculating the impact parameter dependent probability, and subsequently combined with neutron emission probabilities associated with Coulomb excitation of the nuclei. These neutron emission scenarios are modeled using a Poisson statistics, allowing the cross sections to be categorized into experimentally relevant neutron classes: 0n0n, 0nXn, XnXn, and inclusive (0n0n+0nXn+XnXn) production.  To enable a direct comparison with experiments, fiducial phase space restrictions corresponding to the acceptance of each detector are implemented.
\end{abstract}

\section{Introduction}
\label{intro}
The production of an electron-positron pair via photon-photon fusion ($\gamma\gamma\rightarrow e^+e^-$) was first theoretically formulated by Gregory Breit and John A. Wheeler, and is now known as the Breit-Wheeler process \cite{breit1934collision}. While the original formulation of Breit-Wheeler analysis showed that directly achieving high energy photon-photon collisions under laboratory conditions is practically impossible, it has been shown that Lorentz-boosted Coulomb fields of highly charged ions pass each other at ultrarelativistic speeds can be reinterpreted as linearly polarized virtual photon fluxes with predominantly transverse components in the appropriate kinematic region, and that this process can be achieved through the fusion of these photons. Using the formalism developed by Evan James Williams and Carl Friedrich von Weizsäcker, differential flux expressions of the photon spectrum obtained as a result of the quantization of the Coulomb field were derived, and it was shown that the photon-photon fusion cross section can be calculated in the collision environment by evolution of these fluxes \cite{weizs1934flicker,williams1934nature}.
The cross sections of electron-positron pairs are quite large so that it is possible to test Quantum Electrodynamics (QED) for the strong field mainly for the ultraperipheral collisions (UPC). UPCs of heavy ions are described by an impact parameter $b$ greater than the sum of the radius of the colliding nuclei. In this case, nuclei pass each other at impact parameters larger than the sum of their nuclear radii, so they avoid direct overlap between the nuclei. Therefore, the hadronic interactions are eliminated, and electromagnetic interactions become dominant. The intense electromagnetic fields are produced by the high charge numbers of the nuclei ($Z_{Au}=79$, $Z_{Pb}=82$) at the close to speed of light so that electromagnetic interactions and photon induced reactions become the dominant focus. Lorentz contracted Coulomb fields in the relativistic limit $(v/c \rightarrow1)$ produced by an ion scales as $E \sim Ze\gamma/b^2$, increasing linearly with the nuclear charge $Z$, inversely with the square of the impact parameter $b$, and proportionally to the Lorentz factor $\gamma$. In the collider frame, the Lorentz factor $\gamma = \mathcal{O}(10^2)$ at Relativistic Heavy Ion Collider (RHIC) and reaches $\mathcal{O}(10^3\text{-}10^4)$ at Large Hadron Collider (LHC), leading to extremely strong fields enough to excite the vacuum and allow the formation of lepton-antilepton pairs through quantum processes \cite{bertulani2005physics,baltz2008physics,brandenburg2023report,nystrand2007}.

For ultrarelativistic motion, the spatial (along the ion direction) width of the Lorentz contracted field is of the order $ b/\gamma$, and therefore it will last until $\Delta t =b/(\gamma c \beta)$ and produce Fourier frequencies up to $\omega_{\max}=(\Delta t)^{-1}= \gamma c \beta/ b$. Then pair production requires $\hbar \omega_{\max} \gtrsim 2mc^2$, where $2m$ is the produced total pair mass. This condition is easily satisfied for electrons at RHIC energies and remains achievable for muons, whereas tau production is strongly suppressed. At LHC energies, the photon energy spectrum extends into the multi GeV range, allowing copious electron and muon pair production and kinematically permitting tau pair production, albeit with significantly smaller cross sections. Additionally, we suppose that the field's energy density will be high enough. By taking $U=e\phi=eEd=eE(2\hbar /mc)$ and setting the distance $d$ to the typical length scale for the produced pair, that is, twice the Compton wavelength of a single lepton. We expect pair production to be possible if this energy equals the mass of the created pair, $2mc^2$. This results in a critical electric field that can be given as
\begin{equation}
E_{\mathrm{crit}} = \frac{m^2 c^3}{e\hbar}.
\end{equation}
The quadratic dependence on the lepton mass implies different threshold values for each lepton species \cite{gucclu1999electromagnetic}. We list the relevant parameters in Table \ref{tab:leptonprops}. In UPCs with gold and lead, for electrons, the maximum field $E_{\max}$ approaches or exceeds $E_{\mathrm{crit}}$ already at RHIC and significantly surpasses it at the LHC. For muons, RHIC lies near threshold, while the LHC clearly exceeds the critical field. Tau production remains close to threshold even at the LHC and is therefore rare.

\begin{table}
\centering
\caption{Lepton properties relevant for pair production.}
\label{tab:leptonprops}
\begin{tabular}{cccc}
\hline
Lepton & Mass (MeV/$c^2$) &  $\lambdabar_C$ (fm) & $E_{\mathrm{crit}}$ (V/m) \\
\hline
Electron ($e^\pm$) & 0.511 & $386.2$ & $1.32 \times 10^{18}$ \\
Muon ($\mu^\pm$)   & 105.66 & $1.868$ & $5.66 \times 10^{22}$ \\
Tau ($\tau^\pm$)   & 1776.86 & $0.111$ & $1.61 \times 10^{25}$ \\
\hline
\end{tabular}
\end{table}

\begin{figure}
\centering
\includegraphics[width=0.48\textwidth]{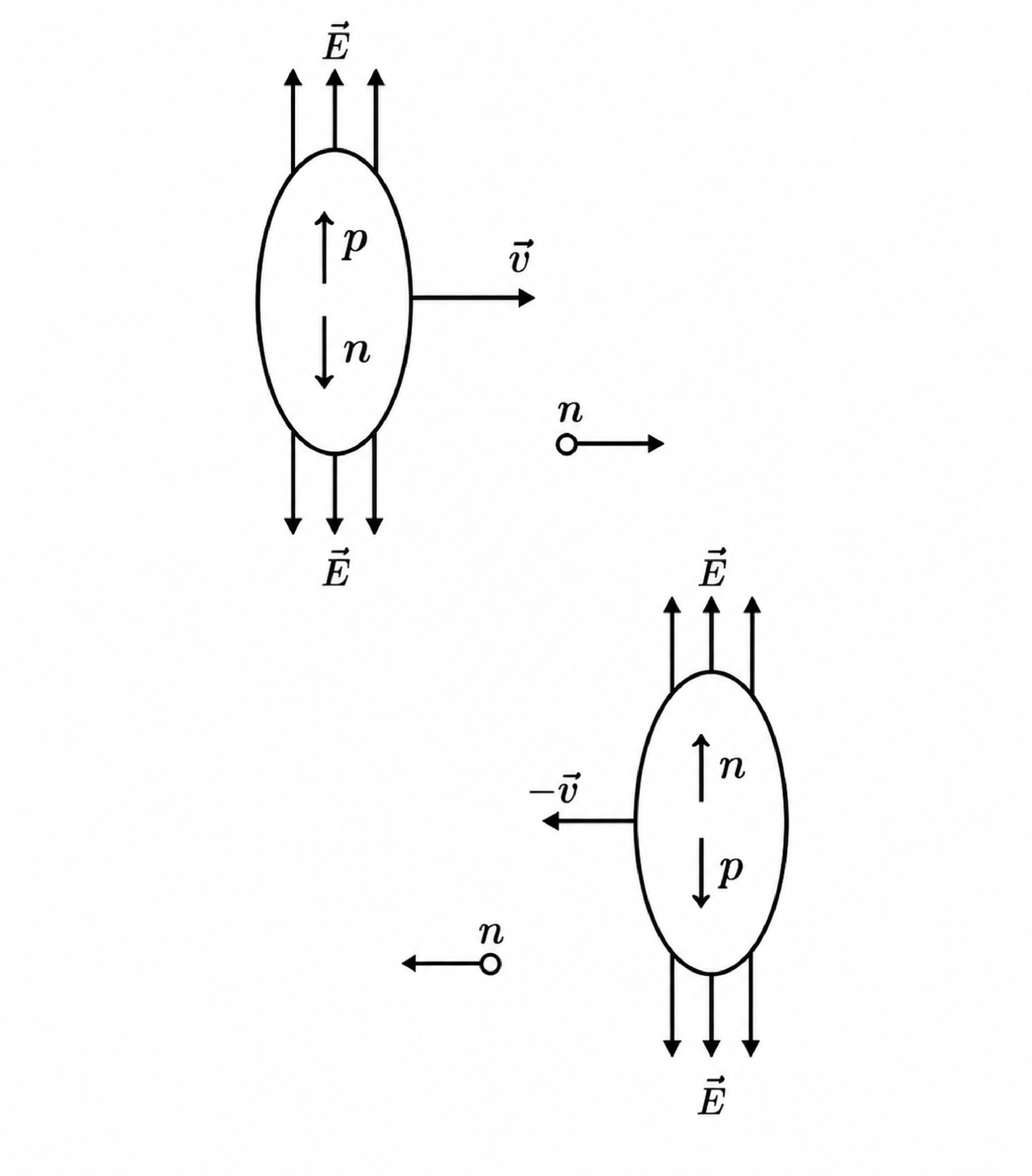}
        \caption{Schematic figure of giant dipole motion in colliding of two relativistic heavy ions. The electric fields of the heavy ions pushes the protons upwards and the neutrons then move downwards. During this motion some neutrons leave the nuclei and move in the beam direction.}
\label{fig:giantdipole}
\end{figure}

These parameters are not only essential for the characterization of the fields, but also play a significant role in describing nuclear excitation processes. In particular, additional photon exchanges between the nuclei can excite one or both ions into higher energy states, which then decay by emitting one or more neutrons. The giant dipole resonance (GDR) is a collective oscillation of protons against neutrons within the nucleus dominates these excitations, while higher resonances provide only subleading contributions at the photon energies relevant for UPCs
\cite{bertulani1988electromagnetic,bertulani1999microscopic,brandenburg2020acoplanarity}. This interaction is depicted in Fig. \ref{fig:giantdipole} in colliding of two relativistic heavy ions.  The electric fields of the heavy ions pushes the protons upwards and the neutrons then move downwards which often results in the emission of a single or multiple neutrons in the beam directions and these neutrons are detected with high precision using Zero Degree Calorimeters (ZDCs).  The classification of events into 0n0n for no neutrons in either ZDC, 0nXn for neutrons detected in one ZDC only, and XnXn for neutrons in both ZDCs. By measuring the fractions of events in each category, the observed dileptons can be corrected for the ion dissociation. Smaller impact parameters, which associate with events containing forward neutrons, increase the photon flux from one or both ions, modifying the invariant mass and rapidity distributions of the lepton pairs. Dilepton photoproduction has been studied by several groups using explicit differential cross-section formulas \cite{Li2020,Klein2020,Wang2021}, while including the GDR excitation has also been widely studied in the literature (see, e.g., Refs. \cite{klein2020photonuclear,harland2023exciting,wang2021lepton,vidovic1993electromagnetic,pshenichnov1999particle,pshenichnov2001mutual,Wang2023}) in particular, our previous studies are presented in Refs. \cite{guclu1995,gucclu2007production} and \cite{csengul2011bound}. 

Recent experimental studies at RHIC and LHC have provided precise measurements of dilepton production in UPCs. These measurements investigate photon - photon interactions in a well defined fiducial phase space and allow for detailed comparisons with theoretical predictions. At RHIC, STAR Collaboration has measured electron - positron pair production in Au+Au collisions at $\sqrt{s_{NN}} = 200~\mathrm{GeV}$, selecting events with single electron pseudorapidity $|\eta^e| < 1$, dielectron rapidity $|y^{ee}| < 1$, transverse momentum $p_T^e > 0.2~\mathrm{GeV}/c$, pair transverse momentum $p_T^{ee} < 0.15~\mathrm{GeV}/c$, and invariant mass in the range $0.4 < M_{ee} < 2.6~\mathrm{GeV}/c^2$ \cite{STAR}. At LHC, ALICE Collaboration has explored low and high mass electron - positron production in Pb+Pb collisions at $\sqrt{s_{NN}} = 2.76~\mathrm{TeV}$. The low mass analysis considered $|\eta^e|, |y^{ee}| < 0.9$ with invariant masses $2.2 < M_{ee} < 2.6~\mathrm{GeV}/c^2$, while the high-mass measurement with invariant masses $3.7 < M_{ee} < 10~\mathrm{GeV}/c^2$ under the same kinematic restrictions \cite{ALICEdielectron}. More recently, ATLAS Collaboration has measured both electron and muon pairs in Pb+Pb collisions at $\sqrt{s_{NN}} = 5.02~\mathrm{TeV}$, with selections $|\eta^e|, |y^{ee}| < 2.5$ for electrons ($p_T^e > 2.5~\mathrm{GeV}/c$, $p_T^{ee} < 2~\mathrm{GeV}/c$, $M_{ee} > 5~\mathrm{GeV}/c^2$) and $|\eta^\mu|, |y^{\mu\mu}| < 2.4$ for muons ($p_T^\mu > 4~\mathrm{GeV}/c$, $p_T^{\mu\mu} < 2~\mathrm{GeV}/c$, $M_{\mu\mu} > 10~\mathrm{GeV}/c^2$) \cite{ATLASdielectron,ATLASdimuon}. Finally, CMS Collaboration has reported the first measurements of tau pairs in Pb+Pb collisions at $\sqrt{s_{NN}} = 5.02~\mathrm{TeV}$ \cite{CMSditau}. For comparison with the CMS measurement, the extrapolated results are considered, whose procedure and implementation will be discussed in detail in Section \ref{sec:10}. The all kinematic restrictions discussed above, summarized in Table~\ref{tab:cross_sections}, define the fiducial phase space for each experiment. 

\begin{table*} 
\centering
\caption{Dilepton measurements at RHIC and LHC with fiducial phase space. In this work, our calculations are based on this experimental restrictions.} \label{tab:cross_sections}
\begin{tabular}{ccccccc}
\hline
System \& Exp. & Process & $|\eta^\ell|$ & $|y^{\ell\ell}|$ & $p_T^\ell$ (GeV/$c$) & $p_T^{\ell\ell}$ (GeV/$c$) & $M_{\ell\ell}$ (GeV/$c^2$) \\ \hline Au+Au 200 GeV, STAR & $\gamma\gamma\!\to e^+e^-$ & $<1$ & $<1$ & $>0.2$ & $<0.15$ & $0.4\!-\!2.6$ \\ Pb+Pb 2.76 TeV, ALICE \\ (low mass) & $\gamma\gamma\!\to e^+e^-$ & $<0.9$ & $<0.9$ & -- & -- & $2.2-2.6$ \\ Pb+Pb 2.76 TeV, ALICE \\ (high mass) & $\gamma\gamma\!\to e^+e^-$ & $<0.9$ & $<0.9$ & -- & -- & $3.7-10$ \\ Pb+Pb 5.02 TeV, ATLAS & $\gamma\gamma\!\to e^+e^-$ & $<2.5$ & $<2.5$ & $>2.5$ & $<2$ & $>5$ \\ Pb+Pb 5.02 TeV, ATLAS & $\gamma\gamma\!\to \mu^+\mu^-$ & $<2.4$ & $<2.4$ & $>4$ & $<2$ & $>10$ \\ Pb+Pb 5.02 TeV, CMS \\(extrapolated) & $\gamma\gamma\!\to \tau^+\tau^-$ & -- & -- & -- & -- & -- \\  
\hline
\end{tabular}
\end{table*}
In this paper, we investigate the production of lepton pairs, starting from QED Feynman diagrams that describe the process $\gamma\gamma \to \ell^+\ell^-$, we take the classical limit for ion trajectories and treat them as external electromagnetic sources. This approach reproduces the Weizsäcker and Williams formalism while in addition to this our method is suitable for detailed numerical implementation, enabling consistent derivation of the impact parameter dependent probability.
\noindent
To evaluate the multidimensional integrals, we use Monte Carlo technique adapted to overcome the rapid oscillation structure caused by Bessel functions. The integration is rearranged so that a nine dimensional integral, becomes a smooth function. For impact parameter $b$, we combine the probability of dilepton production with the probability of nuclear dissociation. The cross section is then obtained by integrating the product of these probabilities over the impact parameter space. This method which accounts for all possible neutron emission configurations injn (i, j=0, 1, X), allowing for realistic modeling of experimentally tagged events. Finally, we calculate differential cross section distributions such as rapidity $|y^{\ell\ell}|$, invariant mass $M_{\ell\ell}$, and transverse momentum $p_T^{\ell\ell}$, and compare findings with recent experimental results to show consistency.

\section{General Formalism}
\label{sec:1}

\begin{figure}
\includegraphics[width=0.48\textwidth]{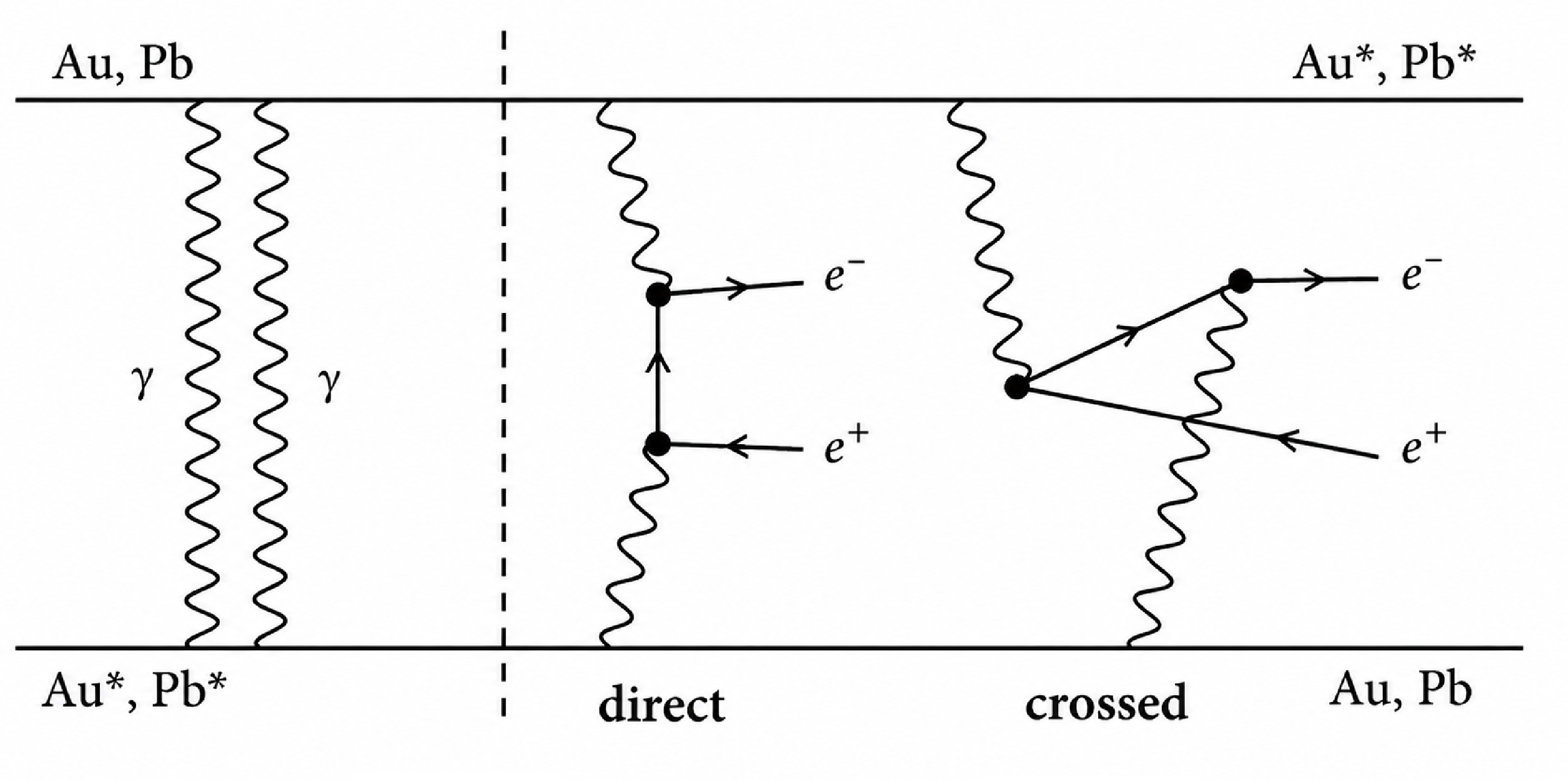}
        \caption{Electron-positron pair production with Coulomb excitation. In this figure only lowest order Feynman diagrams (direct and crossed terms) for electron-positron pair production are shown.}
\label{fig:feynmandiagrams}
\end{figure}

The cross section to produce a lepton-antilepton pair with mutual Coulomb excitation can be written as 
\begin{equation}
	\sigma=\int d^2b \,P_{\ell^+\ell^-}(b)\,[P_{exc}(b)]^2\,P_{nohad}(b),
\end{equation}
where $P_{\ell^+\ell^-}(b)$ is the free pair production probability, $P_{exc}(b)$ is the the probability of a nucleus being excited by a photon from other nucleus. The square of this probability, corresponds to mutual excitation, where both nuclei are excited as shown in Fig. \ref{fig:feynmandiagrams}. $P_{nohad}(b)$ is the probability of no hadronic interaction so that guaranteeing that the collision is ultraperipheral. The task is to determine the each probability accurately. 

\subsection{Free Pair Production Probability}
\label{sec:2}
We begin by writing the cross section of producing electron - positron pair by integrating the $S$ matrix elements over the impact parameter $b$
\begin{equation}
 \sigma_{e^+e^-} = \int d^2b \sum_{k>0} \sum_{q<0} \left| \langle \chi_k^{(+)} | S_{AB} + S_{BA} | \chi_q^{(-)} \rangle \right|^2,
\end{equation}
where the summation over the states $k$ is restricted to those above the Dirac sea, and the summation over the states $q$ is restricted to those occupied in the Dirac sea. Here $S_{AB}$ is the direct term and  $S_{BA}$ is the exchange term of the $S$ matrix elements in Fig. \ref{fig:feynmandiagrams}. 

The result for $S_{AB}$ is
\begin{align}
    \langle \chi_k^{(+)} | S_{AB} | \chi_q^{(-)} \rangle =& \frac{i}{2\beta} \int \frac{d^2 \vec{p}_\perp}{(2\pi)^2} \exp \left[ i \left( \vec{p}_\perp - \frac{\vec{k}_\perp + \vec{q}_\perp}{2} \right) \vec{b} \right] \nonumber\\ 
&* \mathcal{F}(\vec{k}_\perp - \vec{p}_\perp; \omega_A) \mathcal{F}(\vec{p}_\perp - \vec{q}_\perp; \omega_B) \nonumber \\
&*\mathcal{T}_{kq}(\vec{p}_\perp; \beta),
\end{align}
where the functions $\mathcal{F}(q,\omega)$ is the scalar part of the electromagnetic field  of the moving heavy ions in momentum space and $\mathcal{T}_{kq}(p_\perp; \beta)$ is the propagator of the intermediate lepton and the matrix elements for the coupling of the photon to the leptons.
Here, $\omega_A$ and $\omega_B$ are the frequencies of the virtual photons, $\vec{k}_\perp$ and $\vec{q}_\perp$ are momentums of the electrons and positrons. Also, $\ket{X_k^{(+)}}$ refers to the positive-energy spinors, and $\ket{X_q^{(-)}}$ refers to the negative-energy spinors \cite{gucclu1999electromagnetic,csengul2016electromagnetic,sevgik2020multiplescattering}.
Including both the direct and crossed Feynman diagrams, the results for the cross section,
as a function of impact parameter, can be obtained as
\begin{equation}
    \frac{d\sigma}{db} = \int_0^\infty q dq \, b J_0(qb) F(q).    \label{eq:dsigmadb}
\end{equation}
Here, $ F(q)$ is a nine dimensional integral that can be calculated with the Monte Carlo integration method:

\begin{strip}
\begin{align}
   F(q) =& \frac{\pi}{8\beta^2} \sum_{\sigma_k} \sum_{\sigma_q} \int_0^{2\pi} d\phi_q \int \frac{ dk_z dq_z d^2 k_\perp d^2 K d^2 Q} {(2\pi)^{10}}\Bigl\{ \mathcal{F}(\frac{\vec{Q}-\vec{q}}{2}; \omega_A) \mathcal{F}(-\vec{K}; \omega_B)\mathcal{T}_{kq}(\vec{k}_\perp - \frac{\vec{Q}-\vec{q}}{2}; \beta) \nonumber \\ 
   & + \mathcal{F}(\frac{\vec{Q}-\vec{q}}{2}; \omega_A) \mathcal{F}(-\vec{K}; \omega_B) \mathcal{T}_{kq}(\vec{k}_\perp - \vec{K}; -\beta)\Bigl\} \Bigl\{\mathcal{F}(\frac{\vec{Q}+\vec{q}}{2}; \omega_A) \mathcal{F}(-\vec{q}-\vec{K}; \omega_B) \mathcal{T}_{kq}(\vec{k}_\perp - \frac{\vec{Q}+\vec{q}}{2}; \beta)\nonumber\\
    &+ \mathcal{F}(\frac{\vec{Q}+\vec{q}}{2}; \omega_A) \mathcal{F}(-\vec{q}-\vec{K}; \omega_B) \mathcal{T}_{kq}(\vec{k}_\perp +\vec{q}-\vec{K}; -\beta)\Bigl\}.    \label{fqfunction}
\end{align} 
\end{strip}

\begin{figure*}
\centering

\begin{subfigure}[b]{0.32\textwidth}
    \centering
    \includegraphics[width=\textwidth]{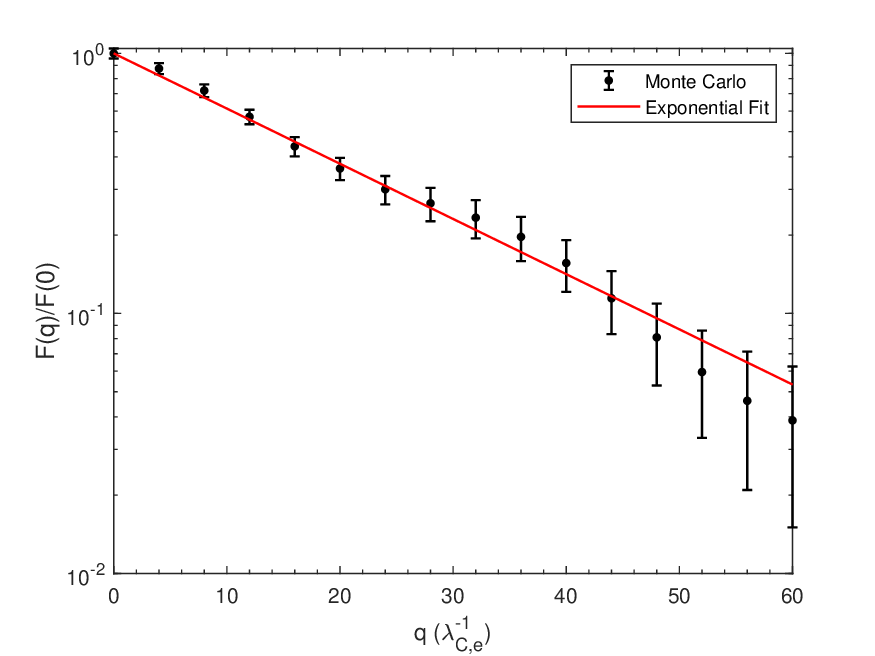}
    \caption{}
    \label{fig:star_e}
\end{subfigure}
\hfill
\begin{subfigure}[b]{0.32\textwidth}
    \centering
    \includegraphics[width=\textwidth]{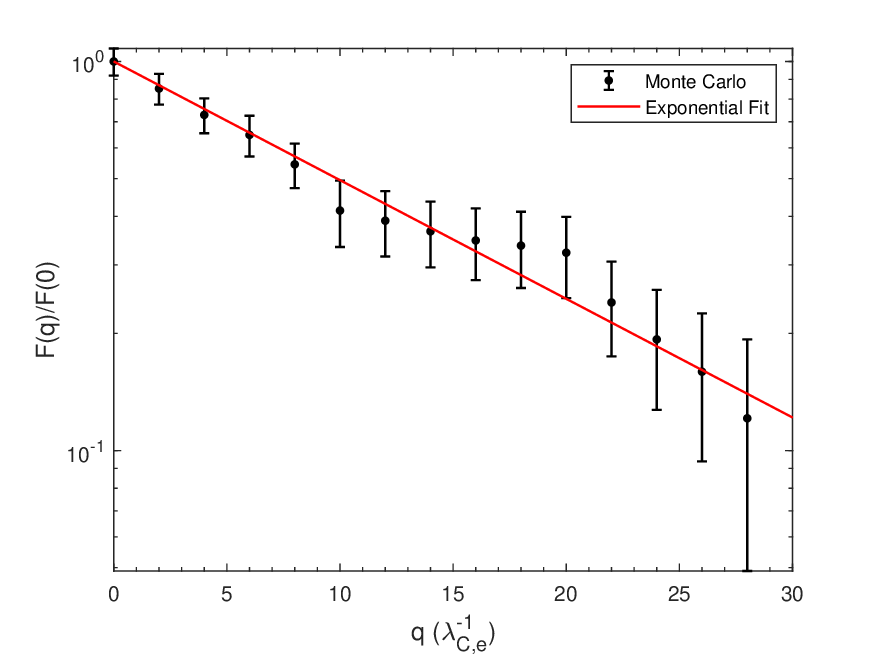}
    \caption{}
    \label{fig:alice_low_e}
\end{subfigure}
\hfill
\begin{subfigure}[b]{0.32\textwidth}
    \centering
    \includegraphics[width=\textwidth]{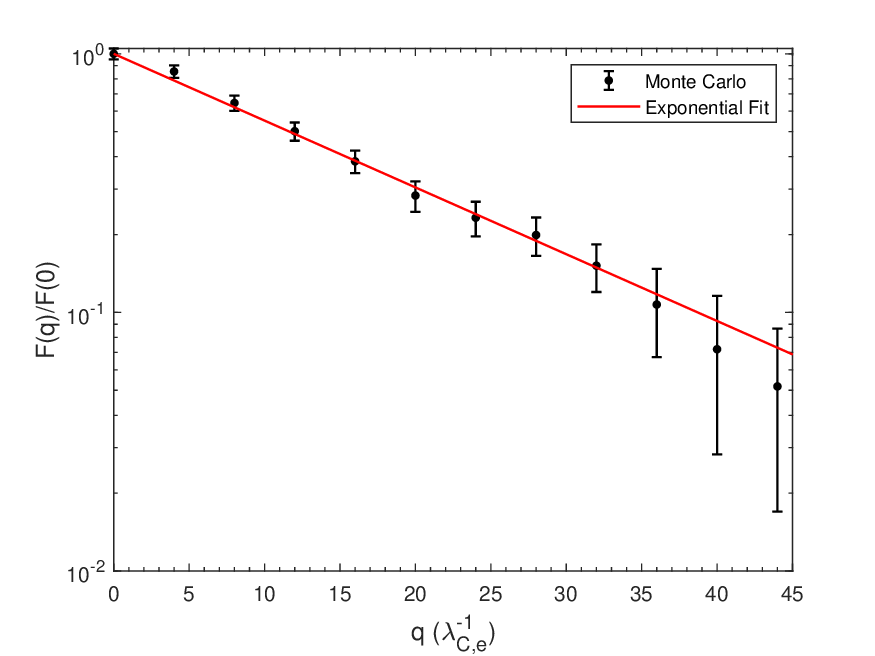}
    \caption{}
    \label{fig:alice_high_e}
\end{subfigure}

\vspace{0.4cm}


\begin{subfigure}[b]{0.32\textwidth}
   
    \includegraphics[width=\textwidth]{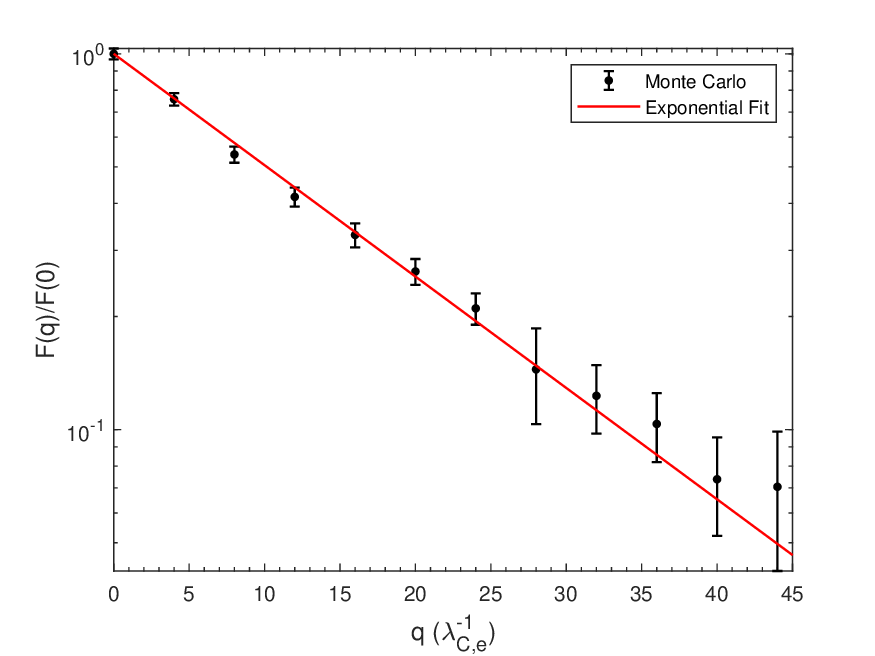}
    \caption{}
    \label{fig:atlas_e}
\end{subfigure}
\hfill
\begin{subfigure}[b]{0.32\textwidth}
    \centering
    \includegraphics[width=\textwidth]{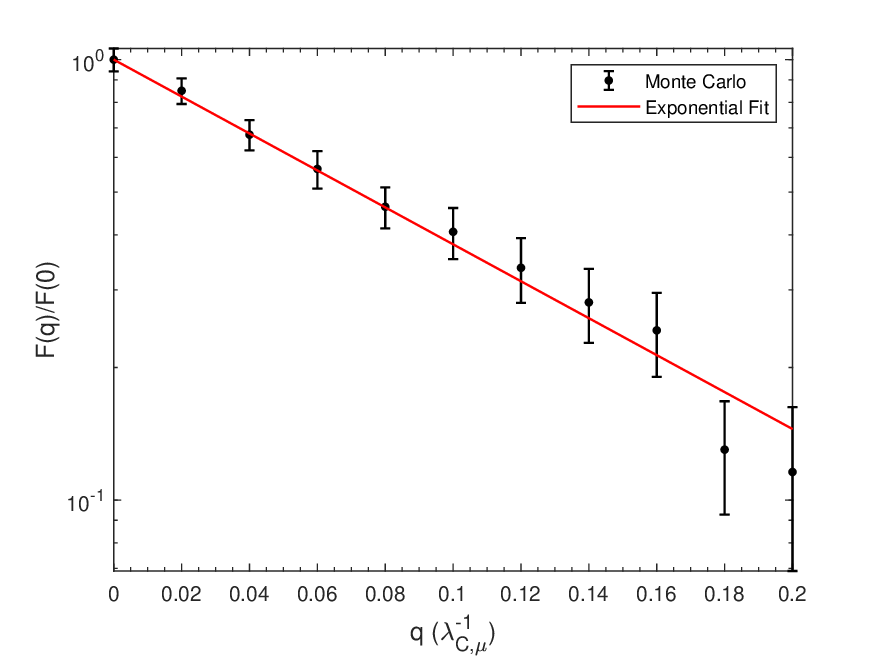}
    \caption{}
    \label{fig:atlas_mu}
\end{subfigure}
\hfill
\begin{subfigure}[b]{0.32\textwidth}
    \centering
    \includegraphics[width=\textwidth]{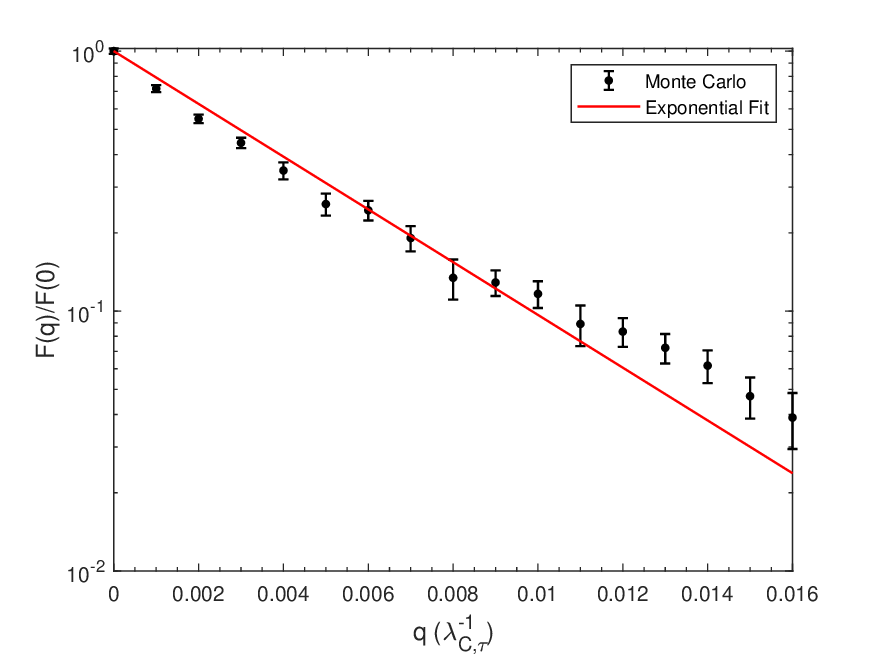}
    \caption{}
    \label{fig:atlas_tau}
\end{subfigure}

\caption{The function F(q)/F(0) is calculated for ultraperipheral heavy ion collisions at RHIC and LHC center-of-mass energies: (a) Electron pair production for STAR at $\sqrt{s_{NN}} = 200$ GeV.
(b) Electron pair production for ALICE at $\sqrt{s_{NN}} = 2.76$ TeV in the low invariant mass region.
(c) Same as (b) but for the high invariant mass region.
(d) Electron pair production for ATLAS at $\sqrt{s_{NN}} = 5.02$ TeV.
(e) Muon pair production for ATLAS at $\sqrt{s_{NN}} = 5.02$ TeV.
(f) Tau pair production for CMS at $\sqrt{s_{NN}} = 5.02$ TeV.  The points show the results of the Monte Carlo calculations for each q value and the smooth curve is our fit for these points. The vertical bars show statistical uncertainties of the data.}
\label{fig:MCfit}
\end{figure*}
A direct numerical evaluation of \eqref{eq:dsigmadb} involves a rapidly oscillating Bessel function $J_0(qb)$, which makes the calculation numerically difficult, particularly at large impact parameters. We found that even an additional transverse momentum integration with up to 100 million Monte Carlo points did not provide satisfactory convergence at large $b$. We therefore introduced a smooth parametrization of the relevant $q$-dependence before performing the final integration, following the strategy previously discussed by \cite{guclu1995}. We have plotted these functions for different heavy ion systems in Fig. \ref{fig:MCfit}. We have done all these calculations for RHIC and LHC energies with applying fiducial
regions matching STAR, ALICE, ATLAS, and CMS. The numerical results show that the probability decreases exponentially as the momentum transfer increases. Since the cross section must vanish in the limit $q\rightarrow\infty$, an exponential function is suitable. Hence, we have found that the most appropriate fit should be in the single exponential form
\begin{equation}
F(q) = F(0)\, e^{-a q},
\label{fq}
\end{equation}
where $F(0)$ corresponds to the total cross section at zero momentum transfer $q=0$ and starting from an initial
parameter set, we determined the unknown parameter $a$ appearing in the exponential term using standard method of nonlinear least-squares (NLLS) curve fitting applied (through software MATLAB) to the Monte Carlo data points (obtained from the Fortran calculations of Eq. \eqref{fqfunction}) where the weights were defined according to the statistical uncertainties of the Monte Carlo results. Meanwhile, we used the Trust - Region algorithm to ensure convergence throughout the fitting process. 

By inserting Eq. \eqref{fq} into the Eq. \eqref{eq:dsigmadb}, the differential cross section can be written as
\begin{equation}
\frac{d\sigma}{db} = F(0)\int_0^\infty dq \, q b \, J_0(qb)\, e^{-a q}.
\end{equation}
Then the probability of lepton - antilepton pair production at a given impact parameter is obtained directly through
\begin{equation}
P_{\ell^+\ell^-}(b) = \frac{1}{2\pi b}\frac{d\sigma}{db}.
\end{equation}

\subsection{Neutron Tagged Probability}
\label{sec:3}

\begin{figure*}
    \centering
    \begin{subfigure}[b]{0.48\linewidth}
    \includegraphics[width=\linewidth]{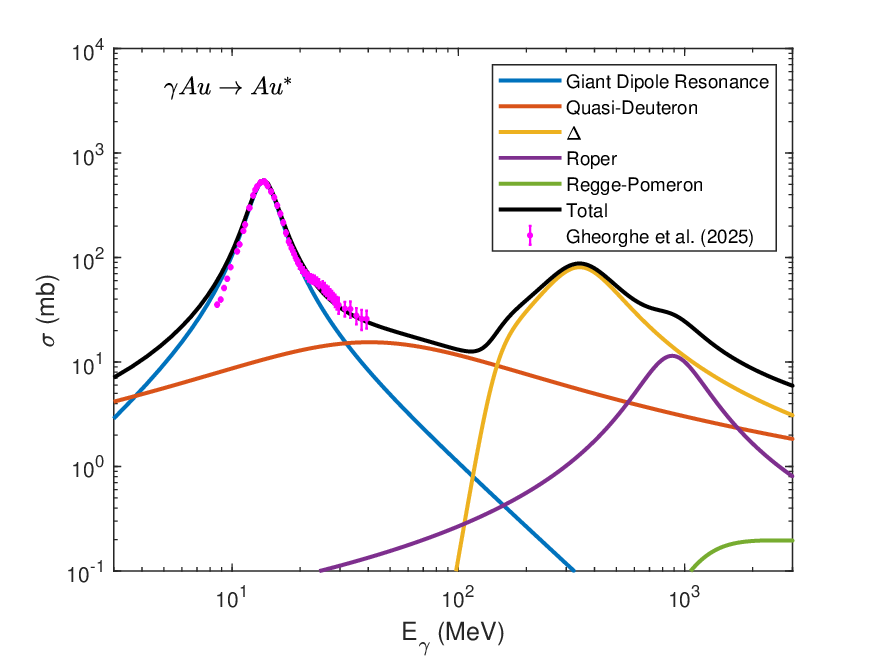}
    \caption{}
    \label{fig:Au_resonance}
\end{subfigure}
\hfill
    \centering
    \begin{subfigure}[b]{0.48\linewidth}
    \includegraphics[width=\linewidth]{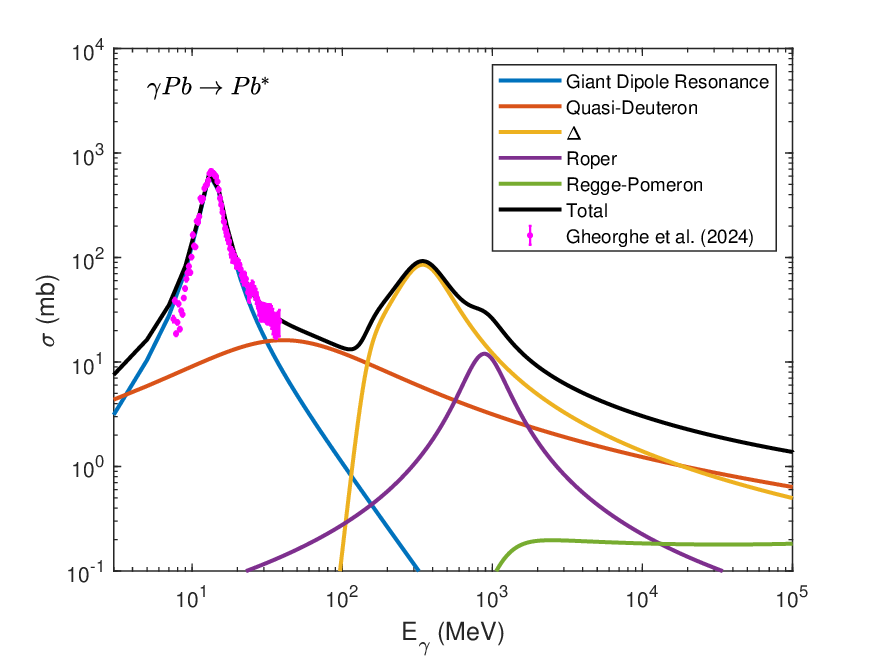}
    \caption{}
    \label{fig:Pb_resonance}
    \end{subfigure}
    \caption{Photoabsorption cross section for the (a) $\gamma Au\rightarrow Au^*$ (b) $\gamma Pb\rightarrow Pb^*$. The GDR component is described by a Lorentzian fit to the experimental photoabsorption data reported by Gheorghe et al. \cite{gheorghegold2025,gheorghelead2024}, while the non-GDR contribution is obtained using the approximation of Kossov \cite{kossov2002approximation,Wellisch:2003wa} }
    \label{fig:Au_Pb_resonances}
\end{figure*}

Following Refs. \cite{klusekgawenda2004,llope1990,pshenichnov2011electromagnetic,bertulani1999microscopic,baltz2009twophoton,hencken1995,baltz1996heavyion}, the mean number of photons absorbed by a nucleus at a given impact parameter is calculated by integrating the flux of photons multiplied by the total photoabsorption cross section over the relevant photon energy range
\begin{align}
P_{\rm exc}(b)
&=\frac{S}{b^2} \nonumber\\
&=\int_{E_{\rm min}}^{E_{\rm max}} dE_\gamma
\frac{1}{E_\gamma}
\frac{Z^2\alpha_{\rm em}}{\pi^2}
\frac{1}{b^2}
\xi^2K_1^2(\xi)
\sigma_{\rm tot}(E_\gamma),
\label{eq:Pexc}
\end{align}
where $K_1$ is the modified Bessel function of the second kind, $\xi=E_\gamma b/\gamma_{lab}$, and $\sigma_{\rm tot}(E_\gamma)$ denotes the total photoabsorption cross section of the nucleus at the photon energy $E_\gamma$. In the ultrarelativistic limit relevant for RHIC and LHC energies, $\xi$ is sufficiently small so that the approximation $\xi^2K_1^2(\xi)\approx1$ can be applied. We can then write \ref{eq:Pexc} as
\begin{equation}
    S=\frac{Z^2\alpha_{\rm em}}{\pi^2}\int_{E_{\rm min}}^{E_{\rm max}} dE_\gamma
\frac{1}{E_\gamma}\sigma_{\rm tot}(E_\gamma).
\label{eq:Svalue}
\end{equation}
The maximum photon energy of a nucleus of radius R and Lorentz boost $\gamma$ is approximately given by $E_{\text{max}}=\gamma\hbar c /R$. For heavy ion collisions, this corresponds to photon energies of up to 3 GeV at RHIC and up to 80-100 GeV at the LHC \cite{ATLASdimuon,baurhencken,steinberg2021ultraperipheral}. We were adopted these values as the upper limits of the photon spectrum when evaluating $S$ values. For the lower integration limit, we choose $E_\text{min}=3$ MeV.

The total photoabsorption cross section is constructed by the contributions from all photon energy regions separately
\begin{equation}
    \sigma_{tot}(E_\gamma)=\sigma_{GDR}(E_\gamma)+\sigma_{non-GDR}(E_\gamma),
    \label{eq:sigtot(E)}
\end{equation}

where $\sigma_{\rm GDR}(E_\gamma)$ represents the contribution from the Giant Dipole Resonance (GDR) region. The GDR dominates the low energy photoabsorption process and for this region we used the standard lorentzian parameterization in Refs.\cite{plujko2018giant,goriely2019reference,klusekgawenda2004} given by
\begin{equation}
  \sigma_{\rm GDR}(E_\gamma)=\sigma_0
\frac{E_\gamma^{2}\Gamma_r}
{\left(E_\gamma^{2}-E_r^{2}\right)^{2}+(E_\gamma\Gamma_r)^2},  
\end{equation}
here we obtained $\sigma_0=\frac{2}{\pi}\sigma_{\rm TRK}S_r$, $\Gamma_r$, and $E_r$ values
by fitting the most recent available data for $^{197}$Au and $^{208}$Pb.

For $^{197}$Au, we used the experimental photoabsorption cross section data reported in 2025 by Gheorghe et al. \cite{gheorghegold2025}, covering the photon energy range from 8.59 MeV to 39.32 MeV. From the fit, we found
$\sigma_0 = 2379$ mb, $E_r = 13.78$ MeV, and $\Gamma_r = 4.42$ MeV.

For $^{208}$Pb, we used the experimental photoabsorption cross section data reported in 2024 by Gheorghe et al. \cite{gheorghelead2024}, covering the photon energy range from 7.5 MeV to 38.02 MeV. From the fit, we found
$\sigma_0 =2367$ mb, $E_r =13.48$ MeV, and $\Gamma_r = 3.953$.

The photonuclear model of Kossov \cite{kossov2002approximation,Wellisch:2003wa} provides approximations for the photoabsorption cross section over the intermediate and high energy photon regions. By using this approximation, the $\sigma_{non-GDR}(E_\gamma)$ is obtained as the sum of the terms
\begin{equation}
\begin{aligned}
\sigma_{\rm non\text{-}GDR}(E_\gamma) =\,
&\sigma_{\rm QD}(E_\gamma)
+\sigma_{\Delta}(E_\gamma) \\
&+\sigma_{\rm Roper}(E_\gamma)
+\sigma_{\rm RP}(E_\gamma),
\end{aligned}
\end{equation}
where $\sigma_{\rm QD}$, $\sigma_{\Delta}$, $\sigma_{\rm Roper}$, and $\sigma_{\rm RP}$ denote the photoabsorption cross sections associated with the quasi-deuteron, $\Delta$ resonance, Roper resonance, and Regge-Pomeron regions, respectively.

In Fig. \ref{fig:Au_resonance} we show the fitted GDR curve together with the Kossov model for $^{197}$Au and in Fig. \ref{fig:Pb_resonance} for $^{197}$Pb. Since the experimental datas of Gheorghe et al. extend into the quasi-deuteron region, we can see from that, the total photoabsorption cross section across the intermediate energy region (the combination of the fitted GDR tail and the quasi-deuteron contribution from the Kossov model)  shows good agreement. This agreement allows us to assume that the adopted model will well established over the entire photon energy range considered. It should also be noted that the $1/{E_\gamma}$ term in Eq.\ref{eq:Svalue} causes the high energy part of the photoabsorption cross section to contribute only weakly to the excitation. 
The S values are now obtained by using Eq. \ref{eq:sigtot(E)} in Eq. \ref{eq:Svalue}. Thus we obtain $S_{Au}=183$ fm$^2$, and $S_{Pb}=223$ fm$^2$.

The quantity $S/b^2$ can be directly employed in the Poisson statistics to determine the probabilities for single and multiple electromagnetic excitations.

Following the prescription of Ref. \cite{baltz2002}, we construct the probabilities for neutron emission based on the Poisson distribution as follows
\begin{align}
P_{(0n)}(b) &= e^{-S/b^2}, \\
P_{(Xn)}(b) &= 1-e^{-S/b^2},
\end{align}
where $P_{(in)}(b)$ with $i=0,X$ denotes the probability corresponding to the neutron emission configurations of no neutron emission and at least one neutron emission, respectively.
We also include mixed neutron emission configurations such as $0nXn$, which are obtained by combining these probabilities.
This formalism provides a consistent description of experimentally relevant neutron emission topologies, labeled as $0n0n$ (no excitation), $0nXn$ (single-sided excitation), and $XnXn$ (mutual excitation). Each neutron configuration is directly related with a specific impact parameter region. In particular, large impact parameters contribute to the $0n0n$ channel, while smaller impact parameters enhance the probability of nuclear excitation and therefore increase the relative contribution of channels containing neutron emission.
After obtaining both the lepton-antilepton pair production probability and using the neutron emission probabilities, we can write the total neutron tagged dilepton production cross section as
\begin{equation} \sigma_{\ell^+\ell^-}^{\text{exc}} = 2\pi \int_{b_{\text{min}}}^{\infty} db \, b \, P_{\ell^+\ell^-}(b) P_{(in)}(b)P_{(jn)}(b). \end{equation}
In this expression, the indices $i$ and $j$ label the neutron emission states of the two colliding nuclei and take values $i,j \in \{0, X\}$. With these definition, the product $P_{(in)}(b)P_{(jn)}(b)$ describes all possible neutron emission topologies in UPCs, so different choices of the indices $(i,j)$ allow us to isolate specific neutron tagged event classes $0n0n$, $0nXn$, and $XnXn$, while summing over all possible configurations yields the inclusive dilepton production cross section
\begin{align} \sigma_{\ell^+\ell^-}^{\text{inc}} =& 2\pi \int_{b_{\text{min}}}^{\infty} db \, b \, P_{\ell^+\ell^-}(b) \nonumber\\
&*\Big[ P_{(0n0n)}(b) + P_{(0nXn)}(b) + P_{(XnXn)}(b) \Big]. 
\label{eq:sig_inc_ll}
\end{align}
Here, $b_{\text{min}}$ denotes the minimum impact parameter for an ultraperipheral heavy ion collision. In our calculations, we take $b_{\text{min}} = R_1 + R_2$, corresponding to the geometrical condition that the colliding ions do not overlap, thereby eliminating hadronic interactions. 

\subsection{No-hadronic Interaction Probability}
\label{sec:4}

\begin{figure}

\begin{subfigure}{0.48\textwidth}
\includegraphics[width=\textwidth]{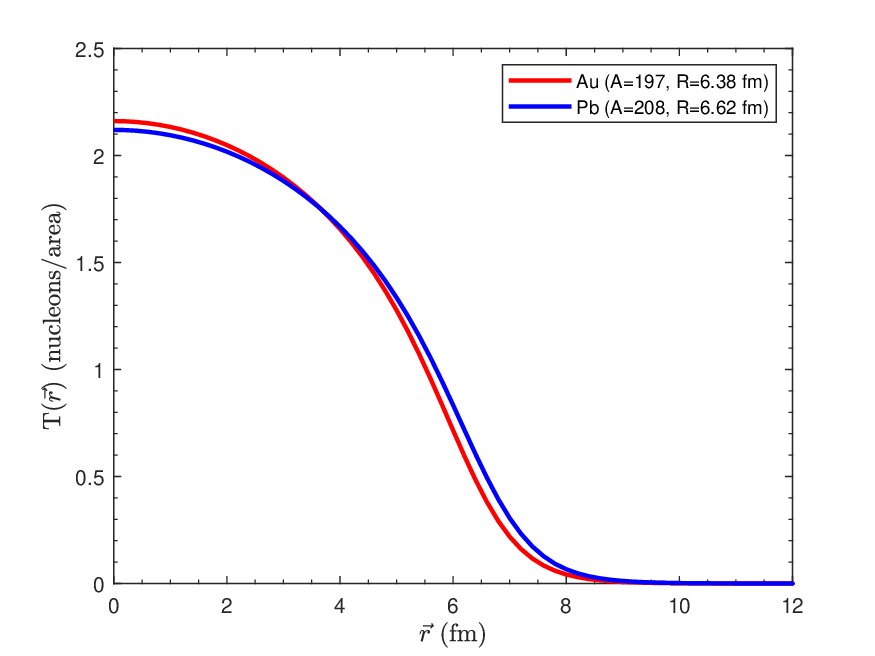}
\caption{}
\label{fig:mnuclearthickness}
\end{subfigure}
\begin{subfigure}{0.48\textwidth}
\includegraphics[width=\textwidth]{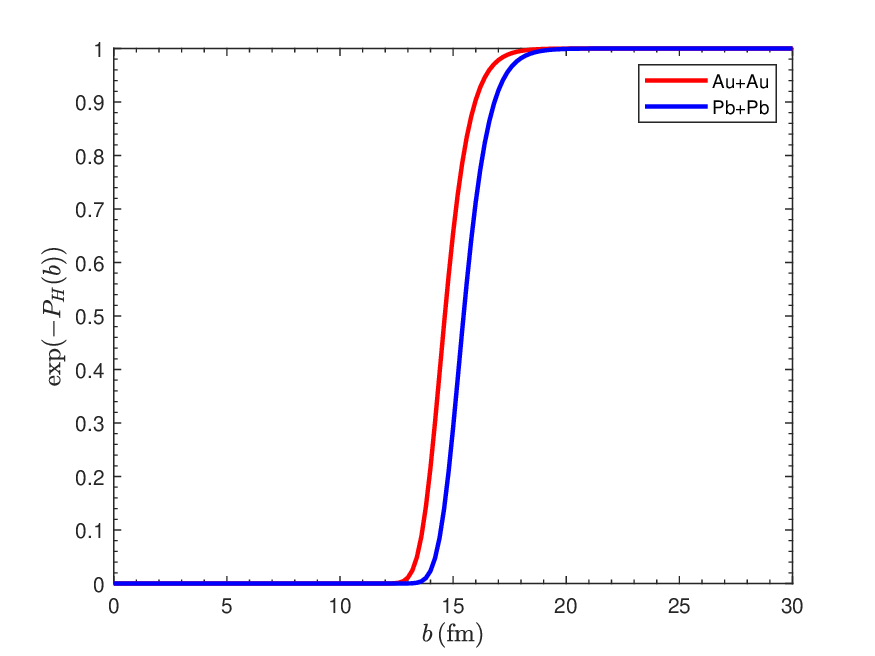}
\caption{}
\label{fig:nohadprobability}
\end{subfigure}

\caption{(a) Comparison of the nuclear thickness functions $T_A(r)$, for Au and Pb nuclei calculated from the corresponding Woods-Saxon density distributions. (b) Probability of no-hadronic interaction $P_{\mathrm{nohad}}(b)$, as a function of the impact parameter for Au+Au and Pb+Pb collisions. The larger radius of the Pb nucleus shifts the transition to $P_{\mathrm{nohad}}(b)=1$ toward larger impact parameters.}
\end{figure}
Although $b_{min}\approx13-14$ fm provides a convenient hard-sphere cutoff for UPCs, it is not sufficiently accurate for the calculation of neutron-tagged (XnXn) cross sections, as it is highly sensitive to the impact parameter near $b=2R$. Instead, a more realistic treatment is adopted in this subsection by calculating the no-hadronic interaction probability. Following Refs. \cite{miller2007glauber,broz2020noon} it is given by 

\begin{equation}
    P_{nohad}(b)=\exp(-P_{had}(b))
    \label{eq:nohadprobability}
\end{equation}
with hadronic interaction probability of the nuclei 
\begin{equation}
    P_{had}(b)=\int d^2r\,
T_A(\vec{r}-\vec{b})
\left[
1-\exp\left(-\sigma_{NN}T_B(\vec{r})\right)
\right],
\end{equation}
where $T$ is the nuclear thickness function and $\sigma_{NN}$ is the inelastic nucleon-nucleon cross section at the corresponding center of mass energy. For Au+Au collisions at RHIC, we use an inelastic nucleon-nucleon cross section of $\sigma_{NN}^{\rm inel}=43.82\pm 0.21$ mb reported in Ref.~\cite{STAR:2020phn}. For Pb+Pb collisions at LHC, we use $\sigma_{NN}^{\rm inel}=70.0\pm1.5$ mb, consistent with the value adopted in the Glauber model calculations of Ref. \cite{CMS:2015nfb}.

The nuclear thickness function is defined as
\begin{equation}
T_A(\vec{r})
=
\int_{-\infty}^{\infty}
\rho_A\!\left(\sqrt{r^2+z^2}\right)\,dz,
\end{equation}
where $\rho_A(r)$ denotes the standard Woods-Saxon parameterization for the nuclear density,
\begin{equation}
\rho_A(r) = \frac{\rho_0}{1 + \exp\left[\frac{r-R}{d}\right]},
\end{equation}
where $r$ is the distance with respect to the nucleus, $R$ is the nuclear radius, $d$ is the skin depth, and $\rho_0$ is the normalization factor so that $\int_0^\infty d^3r \rho_A(r)=A$ nucleons. The upper integration limit of 50 fm is sufficiently large since the Woods-Saxon density becomes negligible beyond this distance. We list the corresponding input parameters used in our calculations in Table \ref{tab:parameters}.

\begin{table}
\caption{The input parameters used  in our calculations.}
\label{tab:parameters}
\begin{tabular}{ccccccc}
\hline\noalign{\smallskip}
Nucleus & $R$ (fm) & $d$ (fm) &$\sigma_{NN}$ (mb)& $N$ & $Z$ & $A$ \\
\hline
$^{197}$Au & 6.380 & 0.535 &43.82& 118 & 79 & 197 \\
$^{208}$Pb & 6.624 & 0.549 &70.0 & 126 & 82 & 208 \\
\hline
\end{tabular}
\end{table}

 Eq. \ref{eq:nohadprobability} is then included as a multiplication in the Eq. \ref{eq:sig_inc_ll} to eliminate contributions from events involving hadronic interactions.

Consequently, this formalism allows us to directly compare our predictions with the STAR measurement \cite{STAR}, since the corresponding data are provided in the XnXn tagged channel. We also provide neutron tagged predictions for other experimentally measured exclusive dilepton productions. In particular, we extend the analysis to heavier lepton final states such as muon and tau pair production.

\section{Results}
\label{sec:5}
\subsection{Dielectron Production at RHIC: Comparison with STAR}
\label{sec:6}
\begin{table*}
\centering
\caption{Neutron tagging of the experimental and calculated results of cross sections for $e^+e^-$ pairs at 200 GeV (STAR)}
\label{tab:neutron_STAR}
\begin{tabular}{ccc}
\hline
Category & Class & Cross section ($\mu$b) \\
\hline

STAR (tagged) \cite{STAR} & XnXn 
& $261 \pm 4_{\rm stat} \pm 13_{\rm sys} $  \\

STARLight (tagged) \cite{klein2017starlight} & XnXn 
& 220  \\

Zha et al. (tagged) \cite{zhaetal} & XnXn 
& 260 \\

Wang et al. (tagged) \cite{wangetal} & XnXn 
& $252 \pm 1.6_{\rm stat}$ \\

\hline
This work (tagged) & 0n0n & 4371 \\
This work (tagged) & 0nXn & 1203  \\
This work (tagged) & XnXn & 274.5  \\

\hline
This work (inclusive) & 0n0n+0nXn+XnXn & 5850 \\
\hline
\end{tabular}
\end{table*}

In this section, we compare our lowest order QED calculation of $\gamma\gamma$ to $e^+e^-$ pairs  with both STAR measurement and other theoretical calculations of ultraperipheral Au+Au collisions at $\sqrt{s_{NN}} = 200$ GeV. 

STAR Collaboration \cite{STAR} measured tagged dielectron production cross section to be $261 \pm 4_{\rm stat} \pm 13_{\rm sys}  $ $\mu$b in the XnXn neutron emission channel, emphasizing that this value corresponds to a fiducial cross section defined within the detector acceptance selected via neutron tagging in the both ZDCs. In particular, the electrons are required to satisfy $p_{T}^{e} > 0.2$ GeV/c and $|\eta^e| < 1$, while the dielectron system is restricted to have an invariant mass in the range $0.4 < m_{ee} < 2.6$ GeV/c$^2$ and pair transverse momentum $p_T^{ee} < 0.15$ GeV/c.  

We calculated the cross sections separately for different neutron emission scenarios: 0n0n, 0nXn, XnXn, and inclusive configuration and we present this results in Table \ref{tab:neutron_STAR}, where they are compared with the STAR measurement, the corresponding prediction from the STARLight Monte Carlo generator \cite{klein2017starlight} and    calculations using the flux-convolved photoabsorption treatment performed
by Zha et al. \cite{zhaetal} and by Wang et al. \cite{wangetal}. As expected, inclusive cross section has the largest value and our result is $\sigma_{inc}=5850$ $\mu$b. In contrast, XnXn calculation leads to a reduction in the total cross section. We found tagged XnXn cross section to be $\sigma_{XnXn}=274 $ $\mu$b and this is 4-5\% above the measured STAR (XnXn) value as well as other compared calculations mentioned above and exceeds the STARLight (XnXn) calculation by 24.5\%. The close difference with STAR indicates a satisfactory level of agreement with the experimental data. Despite the differences in the photoabsorption treatment and the core theoretical approaches, our present results remain consistent with the majority of the published calculations within the expected theoretical uncertainties.

\begin{figure*}

\begin{subfigure}{0.48\linewidth}
\includegraphics[width=\linewidth]{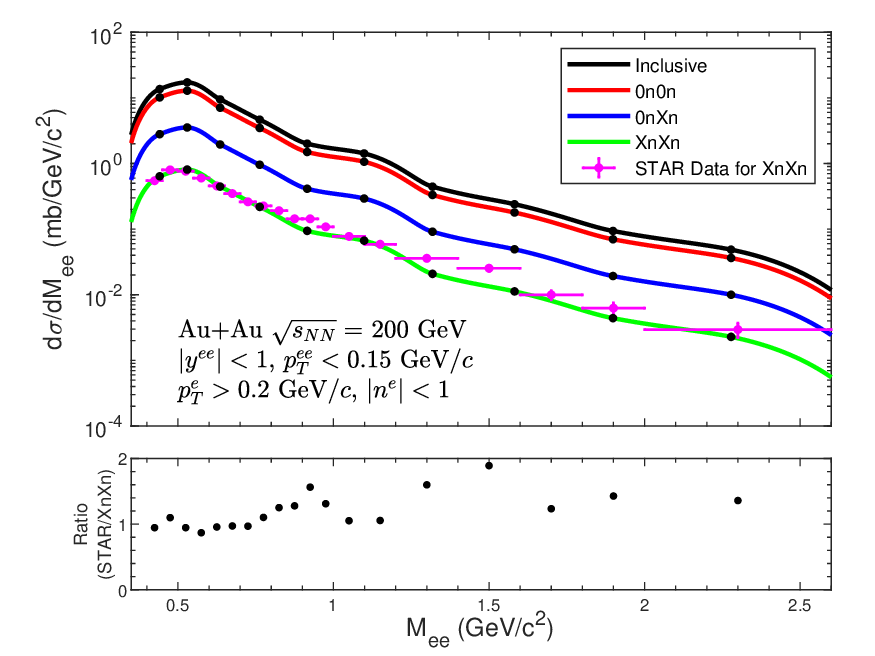}
\caption{}
\label{fig:mass_STAR}
\end{subfigure}
\hfill
\begin{subfigure}{0.48\linewidth}
\includegraphics[width=\linewidth]{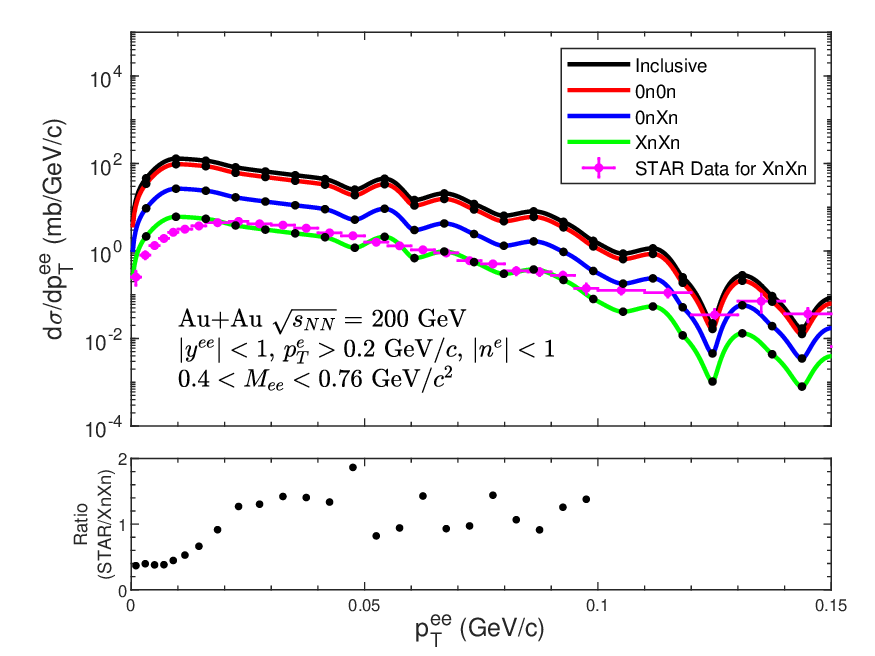}
\caption{}
\label{fig:momentum_STAR}
\end{subfigure}

\caption{The differential cross sections for $e^+e^-$ pairs with respect to (a) the invariant mass $M_{ee}$ (b) the pair transverse momentum $p_T^{ee}$. Results for the inclusive, 0n0n, 0nXn, and XnXn configurations are compared with STAR Data in ultraperipheral Au+Au collisions at $\sqrt{s_{NN}}=$200 GeV for XnXn configuration \cite{STAR}. The statistical uncertainties of the data (vertical bars) and the respective bin sizes (horizontal bars) are shown.}
\end{figure*}

In Fig.  \ref{fig:mass_STAR} we show the differential cross sections as a function of the pair mass $d\sigma/dM_{ee}$ and in Fig. \ref{fig:momentum_STAR} as a function of the transverse momentum $d\sigma/dp_T^{ee}$ for different neutron configurations in the top panel. The data points are taken from STAR measurements for XnXn channel are also plotted for comparison and their ratio is provided in the bottom panel. As seen in the figures, for the invariant mass distribution, our XnXn result is agree well with the measured XnXn data. For the transverse momentum distribution, our XnXn result is consistent with the XnXn data in the  $p_T<0.1$ GeV$/c$ region, while it falls below at high $p_T$ region. Such deviations in the high $p_T$ region are not unexpected, since it should be noted that the differential cross section distributions were obtained directly and $F(q)$ was not recomputed for per kinematic bins. Consequently, the $b$-dependence of the pair kinematics (harder photon $k_\perp$ at small $b$) is lost within each neutron class.

\subsection{Dielectron Production at LHC: Comparison with ALICE}
\label{sec:7}

In this section, we compare our $\gamma \gamma \rightarrow e^+e^-$ production cross section results with the experimental measurements reported by ALICE Collaboration at 2.76 TeV, as well as with the result from the STARLight Monte Carlo generator. 

ALICE Collaboration \cite{ALICEdielectron} measured the two-photon dielectron production inclusive cross section in two separate invariant mass regions, $2.2<M_{ee}<2.7$ and $3.7<M_{ee}<10$ GeV/c$^2$ with electrons are required to satisfy pseudorapidity cuts of $|\eta^e| < 0.9$, obtaining values of 154 $\pm 11_{\rm stat}$ $\mu$b and 91 $\pm 10_{\rm stat}$ $\mu$b, respectively. The separation into two mass intervals is motivated by the need to avoid contamination from vector meson resonances, in particular the $J/\psi$ peak around 3.1 GeV. By selecting regions below and above the resonance, the measurement isolates the continuum $\gamma\gamma\rightarrow e^+e^-$ process, allowing a direct comparison with lowest order QED predictions. 

 \begin{table*}
\centering
\caption{Neutron tagging of the experimental and calculated results of cross sections for $e^+e^-$ pair at 2.76 TeV (ALICE) for low and high mass regions.}
\label{tab:neutron_ALICE_combined}
\begin{tabular}{cccc}
\hline \noalign{\smallskip}
Category & Class  & Low mass ($\mu$b) & High mass ($\mu$b) \\
\hline

ALICE (inclusive) \cite{ALICEdielectron} & 0n0n+0nXn+XnXn 
& 154 $\pm 11_{\rm stat}$ 
& 91 $\pm 10_{\rm stat}$ \\

 STARLight (inclusive) \cite{klein2017starlight}& 0n0n+0nXn+XnXn 
& 128 
& 77 \\

\hline
This work (tagged)& 0n0n & 114 & 66.1 \\
This work (tagged)& 0nXn & 26.9 & 17.4 \\
This work (tagged)& XnXn & 6.06 & 4.15 \\

\hline
This work (inclusive) & 0n0n+0nXn+XnXn 
& 147 
& 87.7 \\
\hline
\end{tabular}
\end{table*}

\begin{figure*}
\centering
\begin{subfigure}{0.45\linewidth}
\centering
\includegraphics[width=\linewidth]{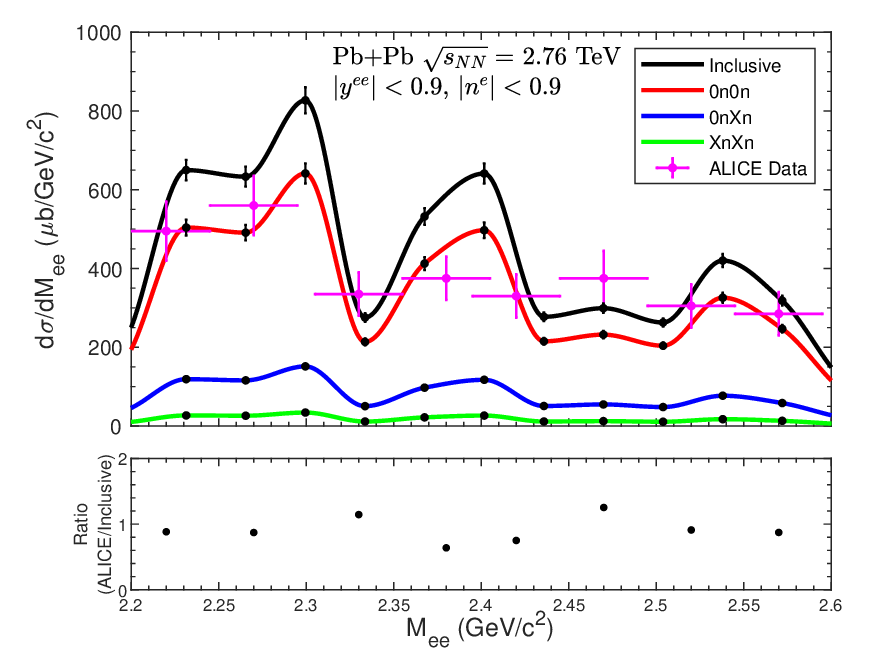}
\caption{}
\label{fig:lowmass_ALICE}
\end{subfigure}
\hspace{0.05\linewidth}
\begin{subfigure}{0.45\linewidth}
\centering
\includegraphics[width=\linewidth]{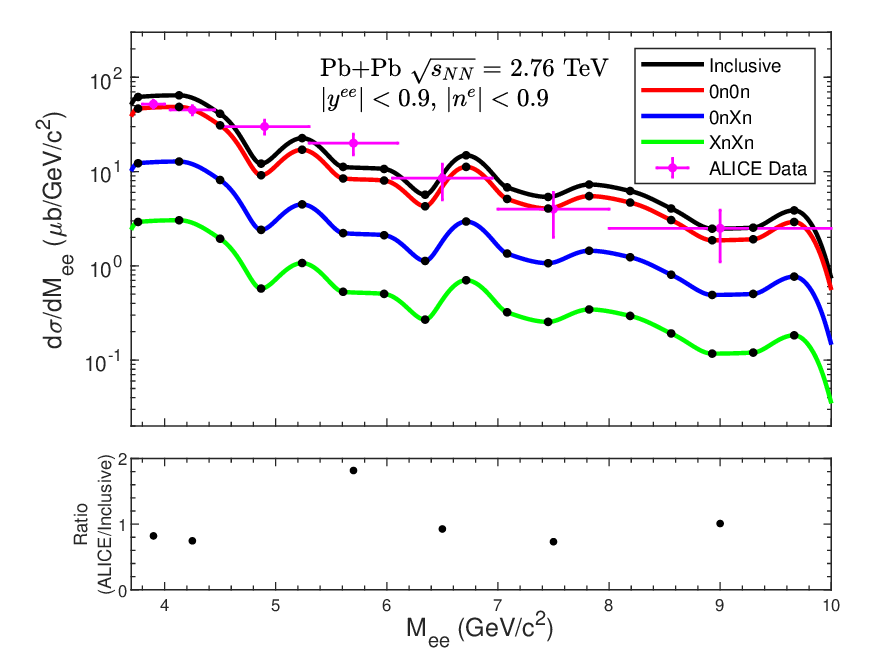}
\caption{}
\label{fig:highmass_ALICE}
\end{subfigure}
\caption{The differential cross sections for $e^+e^-$ pairs with respect to (a) the low invariant mass $2.2<M_{ee}<2.6$ GeV/$c^2$ (b) the high invariant mass $3.7<M_{ee}<10$ GeV/$c^2$. Results for the inclusive, 0n0n, 0nXn, and XnXn configurations are
compared with ALICE Data in ultraperipheral Pb+Pb collisions at $\sqrt{s_{NN}}=$2.76 TeV \cite{ALICEdielectron}. The statistical uncertainties of the data (vertical bars) and the respective bin sizes (horizontal bars) are shown.}
\label{fig:masses}

\end{figure*}
We calculated the cross sections for different neutron emission scenarios: 0n0n,
0nXn, XnXn, and inclusive configuration for low mass region and high mass region separately and we present this results in Table \ref{tab:neutron_ALICE_combined}, together with the STARLight prediction and the ALICE measurements for both low mass and high mass regions.

A key observation is that we found the inclusive cross section $\sigma_{inc}=147$ $\mu$b for low mass region and $\sigma_{inc}=87.7$ $\mu$b for the high mass region and these are 3-5\% below the ALICE inclusive measurement but 13-15\% above the STARLight inclusive calculation for both mass regions. The very close difference with the ALICE results demonstrate the consistency of our calculation with the experimental data.

On the other hand, the 0n0n channel contributes with cross sections of 114 (66.1) $\mu$b in the low (high) mass region. The 0nXn and XnXn channels contribute 26.9 (17.4) $\mu$b and 6.06 (4.15) $\mu$b, respectively.

 Furthermore, we show the differential cross sections as a function of the pair mass $d\sigma/dM_{ee}$ for low mass region in Fig. \ref{fig:lowmass_ALICE} and for high mass region in Fig. \ref{fig:highmass_ALICE} with different neutron configurations in the top panel. The data points are taken from ALICE measurements for inclusive configuration are also plotted for comparison and their ratio is provided in the bottom
panel. The apparent fluctuations in the top panels are further emphasized by the use of a linear vertical representation rather than a logarithmic scale and therefore do not reflect a significant discrepancy between the results. The data to theory ratios remain close to unity over the entire investigated mass ranges, indicating an overall good agreement between our calculations and the experimental results of ALICE within the considered kinematic cuts. 

\subsection{Dielectron Production at LHC: Comparison with ATLAS}
\label{sec:8}
\begin{table*}
\centering
\caption{Neutron tagging of the experimental and calculated results of cross sections for $e^+e^-$ pair at 5.02 TeV (ATLAS).}
\label{tab:neutron_ATLAS}
\begin{tabular}{ccc}
\hline\noalign{\smallskip}
Category & Class  & Cross section ($\mu$b) \\
\hline

ATLAS (inclusive) \cite{ATLASdielectron}& 0n0n+0nXn+XnXn 
& 215 $\pm 1_{\rm stat} \pm 4$  \\

STARLight (inclusive) \cite{klein2017starlight} &  0n0n+0nXn+XnXn
& 196.9  \\
 
SuperChic 3 (inclusive) \cite{SuperChic}&  0n0n+0nXn+XnXn
& 235.1  \\

\hline
This work (tagged)& 0n0n & 180  \\
This work (tagged) & 0nXn & 43.3  \\
This work (tagged) & XnXn & 9.83  \\

\hline
This work (inclusive) & 0n0n+0nXn+XnXn & 233 \\
\hline
\end{tabular}
\end{table*}

\begin{figure*}
\centering
\begin{subfigure}{0.48\linewidth}
\centering
\includegraphics[width=\linewidth]{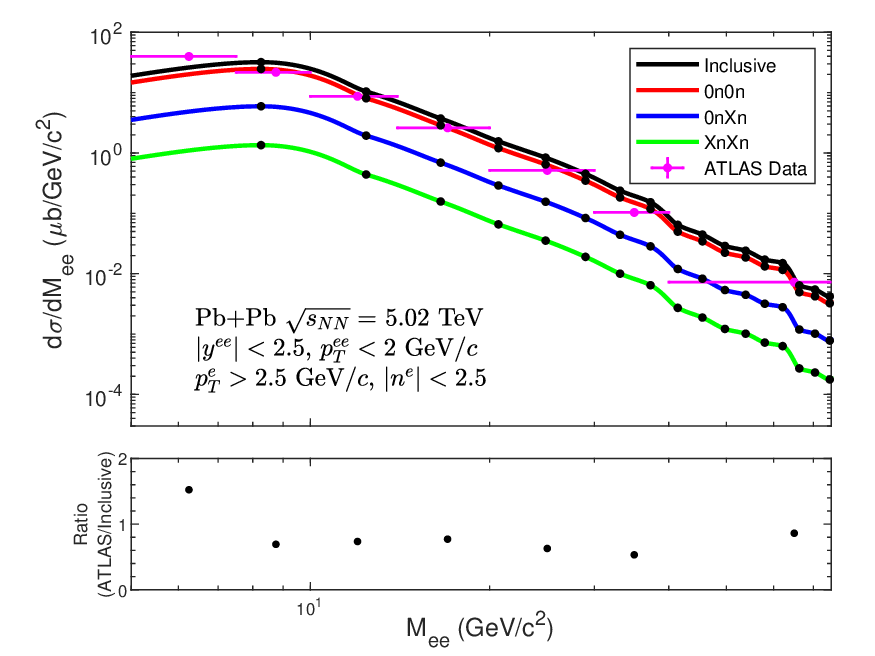}
\caption{}
\label{fig:atlas_mass}
\end{subfigure}
\hfill
\begin{subfigure}{0.48\linewidth}
\centering
\includegraphics[width=\linewidth]{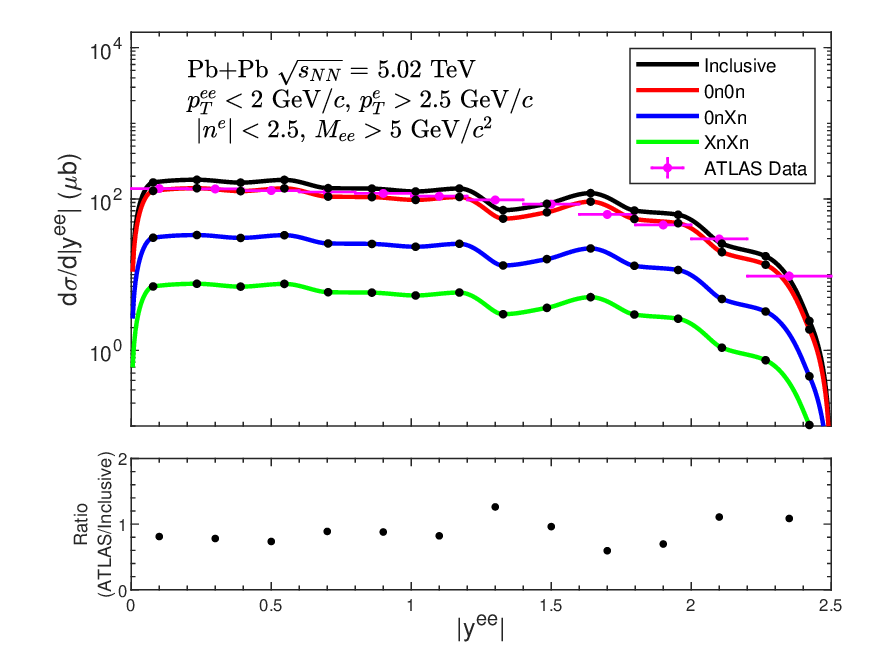}
\caption{}
\label{fig:atlas_rapidity}
\end{subfigure}

\caption{The differential cross sections for $e^+e^-$ pairs with respect to (a) the invariant mass $M_{ee}$ (b) the pair rapidity $|y^{ee}|$. Results for the inclusive, 0n0n, 0nXn, and XnXn configurations are compared with ATLAS Data in ultraperipheral Pb+Pb collisions at $\sqrt{s_{NN}}=$5.02 TeV \cite{ATLASdielectron}. The statistical uncertainties of the data (vertical bars) and the respective bin sizes (horizontal bars) are shown.}
\label{fig:atlas_distributions}

\end{figure*}

In this section, we compare our lowest order QED calculation of $\gamma\gamma$ to $e^+e^-$ pairs  with ATLAS measurement and other calculations of ultraperipheral Pb+Pb collisions at $\sqrt{s_{NN}} = 5.02$ TeV. The fiducial phase space considered in this analysis follows the experimental definition, requiring $p_T^e > 2.5$ GeV/c, $|\eta^e| < 2.5$, dilepton invariant mass $M_{ee} > 5$ GeV/c$^2$, and pair transverse momentum $p_T^{ee} < 2$ GeV/c. Within this kinematic region, ATLAS Collaboration \cite{ATLASdielectron} measured fiducial dielectron production inclusive cross section to be $215 \pm 1_{\rm stat} \pm 4_{\rm sys}\ \mu\mathrm{b}$. 

 We calculated the cross sections separately for different neutron emission scenarios: 0n0n, 0nXn, XnXn, and inclusive case and we present this results in Table \ref{tab:neutron_ATLAS}, where they are compared with the ATLAS measurement and the corresponding predictions from the STARLight Monte Carlo generator \cite{klein2017starlight} and SuperChic 3 \cite{SuperChic}. Our inclusive cross section result is $\sigma_{inc}=233$ $\mu$b and this is 8.3\% above the measured ATLAS inclusive value, 18.3\% above the STARLight calculation, 0.89\% below the SuperChic 3 calculation.  

 \begin{figure*}
\centering
\begin{subfigure}{0.32\linewidth}
\centering
\includegraphics[width=\linewidth]{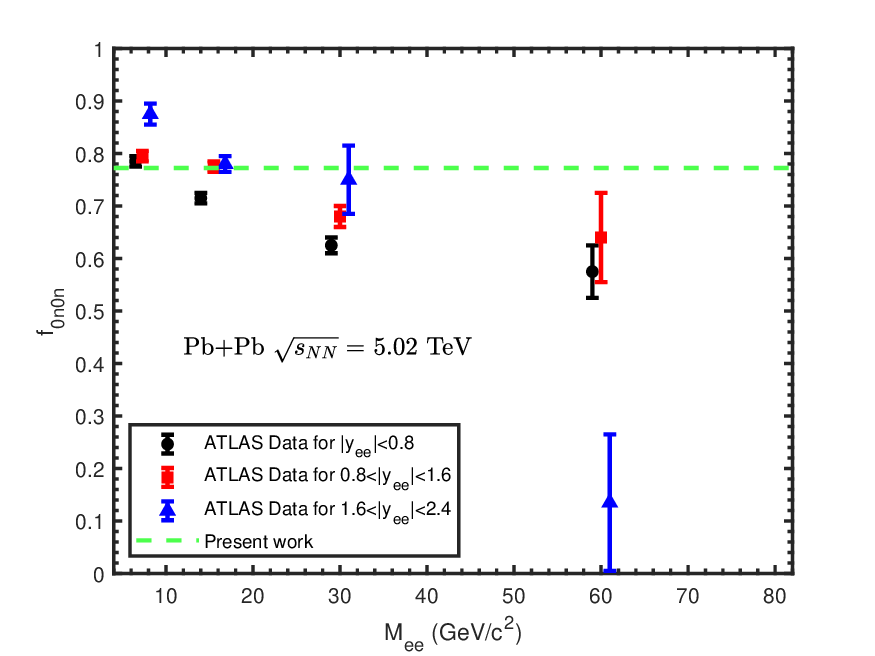}
\caption{}
\label{fig:frac1_atlas_electron}
\end{subfigure}
\hfill
\begin{subfigure}{0.32\linewidth}
\centering
\includegraphics[width=\linewidth]{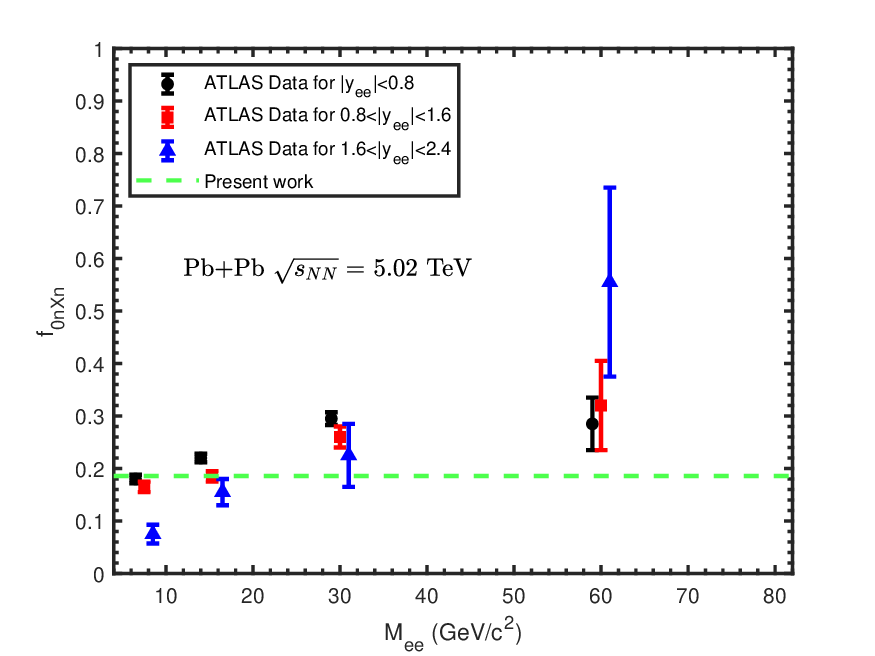}
\caption{}
\label{fig:frac2_atlas_electron}
\end{subfigure}
\hfill
\begin{subfigure}{0.32\linewidth}
\centering
\includegraphics[width=\linewidth]{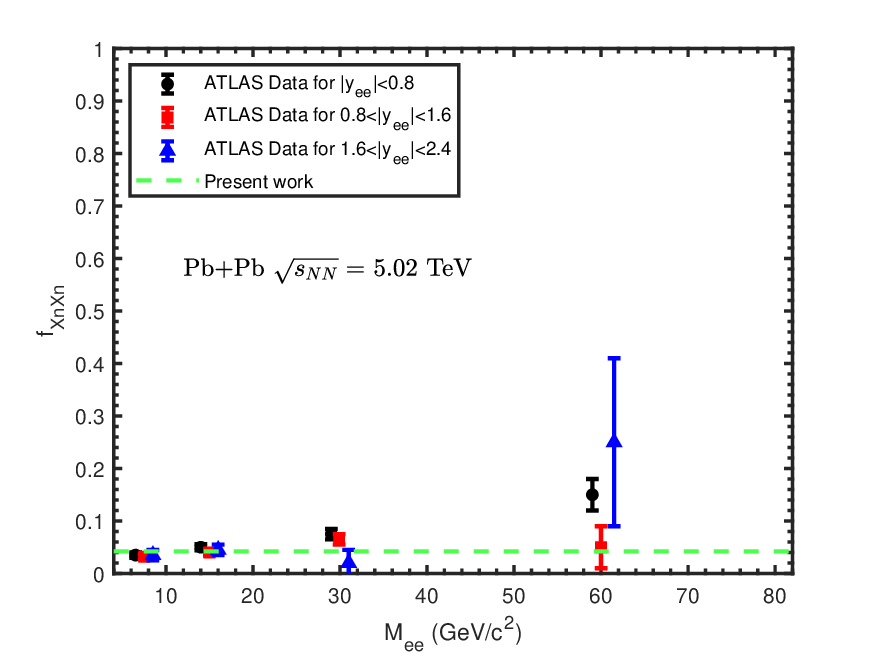}
\caption{}
\label{fig:frac3_atlas_electron}
\end{subfigure}

\caption{Fractions of the (a) 0n0n (b) 0nXn (c) XnXn categories as a function of $M_{ee}$ in three bins of $|y_{ee}|$. The experimental data and their uncertainties were reproduced from the measurements reported by the ATLAS \cite{ATLASdielectron}. Error bars represent statistical uncertainties. The dashed line shows the present work. Points for $|y_{ee}|<0.8$ and $1.6 < |y_{ee}| < 2.4$ are displaced horizontally for better visibility.}
\label{fig:fracs_electron}

\end{figure*}

By decomposing the total cross section into neutron emission channels, the dominant contribution arises from the 0n0n class, with a value of 180 $\mu$b. The 0nXn and XnXn channels, associated with single and mutual electromagnetic dissociation, contribute $43.3\ \mu\mathrm{b}$ and $9.83\ \mu\mathrm{b}$, respectively. 

In Fig. \ref{fig:atlas_distributions} we show the differential cross sections as functions of invariant mass $M_{ee}$ and rapidity $|y^{ee}|$, for inclusive and neutron-tagged classes in the top panel. The data points are taken from ATLAS measurements for inclusive channel are also plotted for comparison and their ratio is provided in the bottom panel. The vertical and horizontal axes are plotted on a logarithmic scale. As shown in the figures, the calculated inclusive differential distributions are in good agreement with the corresponding measurements of ATLAS. The data to theory ratios stay close to unity throughout the considered invariant mass range, and especially for the pair rapidity, present theory provide a satisfying description of dielectron production in UPCs within the
considered kinematic cuts.

In Fig. \ref{fig:fracs_electron}, the fractions of the 0n0n, 0nXn, and XnXn neutron emission categories are compared with the ATLAS measurements. It is observed that the present work is in good agreement with the data in the low invariant mass region for all fractions. At higher invariant masses, however, the calculated 0n0n fraction tends to lie slightly above the experimental values, whereas the predicted 0nXn and XnXn fractions remain somewhat below the measurements. This behavior originates from the simplified treatment adopted in the present calculation. We were calculated the fractions only from the ratios of the corresponding cross sections  $\sigma_{\mathrm{0n0n}}/\sigma_{\mathrm{inc}}$, $\sigma_{\mathrm{0nXn}}/\sigma_{\mathrm{inc}}$ and $\sigma_{\mathrm{XnXn}}/\sigma_{\mathrm{inc}}$, without performing a convolution with the photon flux. Consequently, the calculated fractions become independent of the kinematic variables and appear as constant horizontal lines across all kinematic bins.

\subsection{Dimuon Production at LHC: Comparison with ATLAS}
\label{sec:9}

\begin{table*}
\centering
\caption{Neutron tagging of the experimental and calculated results of cross sections for $\mu^+\mu^-$ pair at 5.02 TeV (ATLAS).}
\label{tab:neutron_ATLAS_muon}
\begin{tabular}{ccc}
\hline\noalign{\smallskip}
Category & Class & Cross section ($\mu$b) \\
\hline

ATLAS (inclusive) \cite{ATLASdimuon} & 0n0n+0nXn+XnXn
& 34.1 $\pm 0.3_{\rm stat} \pm 0.7_{\rm sys}$  \\

STARLight (inclusive) \cite{klein2017starlight} & 0n0n+0nXn+XnXn 
& 32.1  \\
 
 STARLight+Pythia8 (inclusive) \cite{pythia} &  0n0n+0nXn+XnXn
& 30.8  \\

\hline
This work (tagged)& 0n0n & 26.1  \\
This work (tagged) & 0nXn & 8.21  \\
This work (tagged) & XnXn & 2.15  \\

\hline
This work (inclusive) & 0n0n+0nXn+XnXn & 36.5  \\
\hline
\end{tabular}
\end{table*}

\begin{figure*}
\centering
\begin{subfigure}{0.32\textwidth}
\centering
\includegraphics[width=\textwidth]{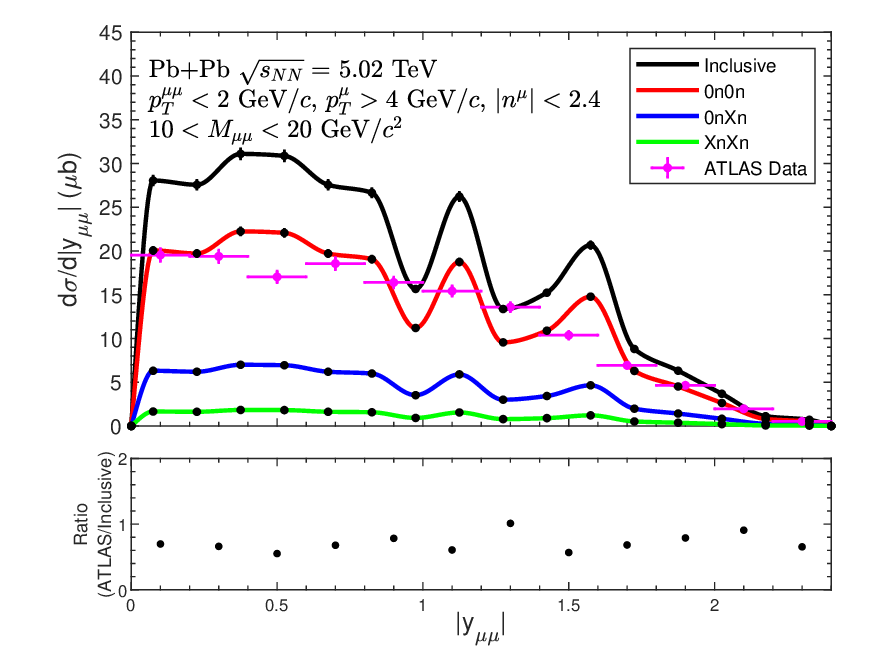}
\caption{}
\label{fig:rapidity_10_20}
\end{subfigure}
\hfill
\begin{subfigure}{0.32\linewidth}
\centering
\includegraphics[width=\textwidth]{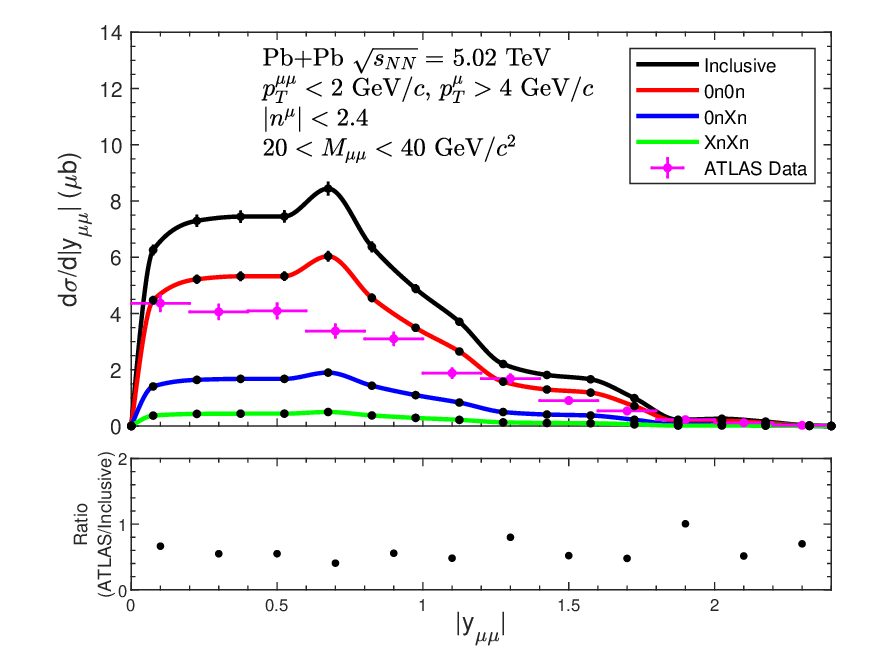}
\caption{}
\label{fig:rapidity_20_40}
\end{subfigure}
\hfill
\begin{subfigure}{0.32\textwidth}
\centering
\includegraphics[width=\textwidth]{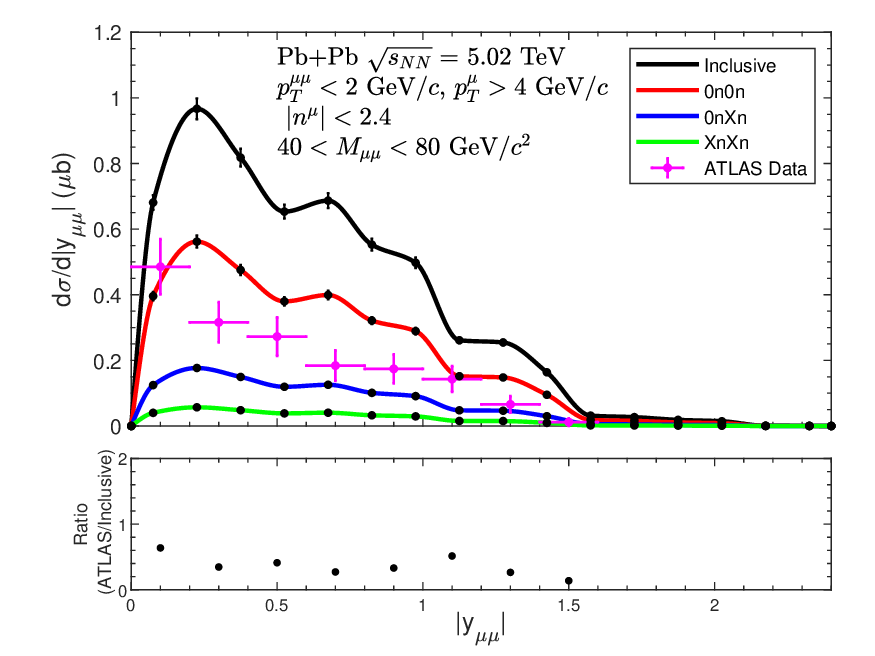}
\caption{}
\label{fig:rapidity_40_80}
\end{subfigure}

\caption{The differential cross sections for $\mu^+\mu^-$ pairs with respect to the pair rapidity $|y_{\mu\mu}|$ in bins of (a) $10<M_{\mu\mu}<20$ (b) $20<M_{\mu\mu}<40$ (c) $40<M_{\mu\mu}<80$.  Results for the inclusive, 0n0n, 0nXn, and XnXn configurations are compared with ATLAS Data in ultraperipheral Pb+Pb collisions at $\sqrt{s_{NN}}=$5.02 TeV \cite{ATLASdimuon}. The statistical uncertainties of the data (vertical bars) and the respective bin sizes (horizontal bars) are shown.}
\label{fig:rapidity_mass_bins}

\end{figure*}

\begin{figure*}
\begin{subfigure}{0.32\linewidth}
\centering
\includegraphics[width=\linewidth]{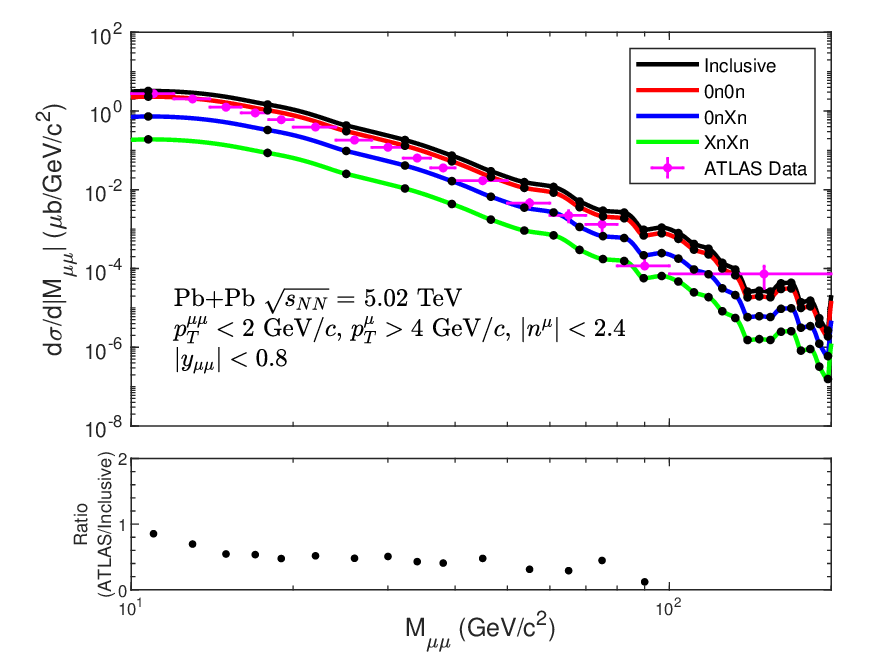}
\caption{}
\label{fig:mass_y08}
\end{subfigure}
\hfill
\begin{subfigure}{0.32\linewidth}
\centering
\includegraphics[width=\linewidth]{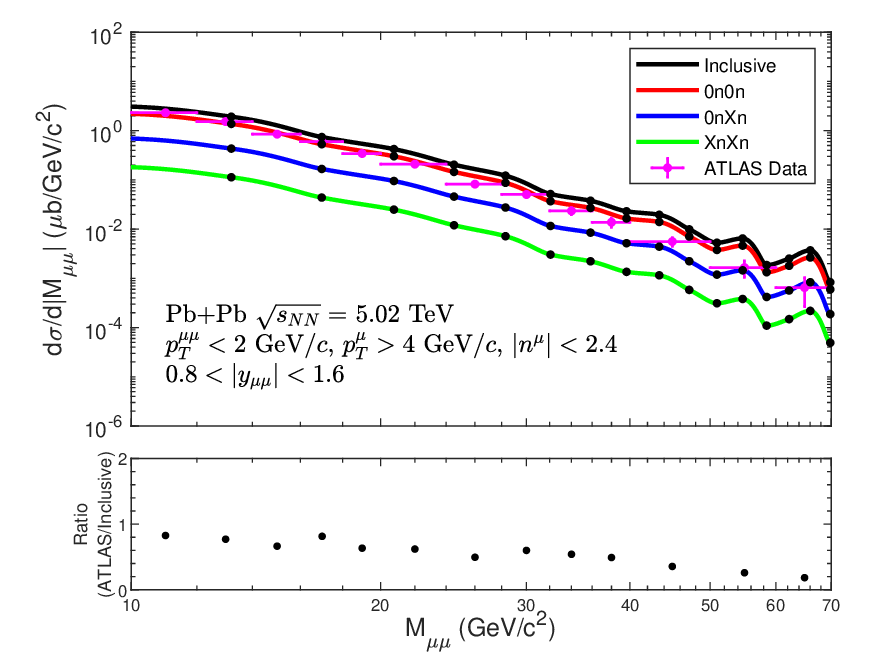}
\caption{}
\label{fig:mass_08y16}
\end{subfigure}
\hfill
\begin{subfigure}{0.32\linewidth}
\centering
\includegraphics[width=\linewidth]{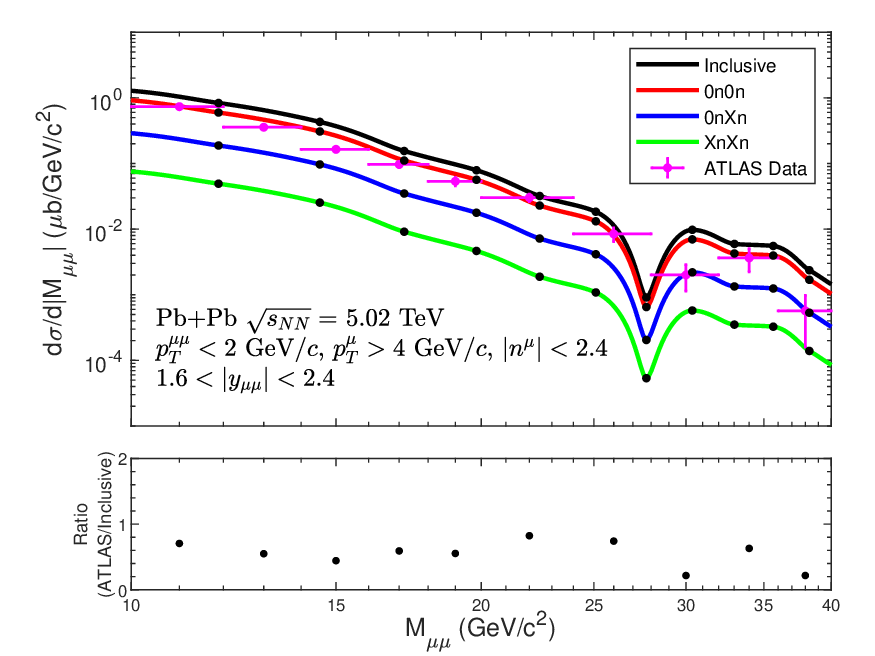}
\caption{}
\label{fig:mass_16y24}
\end{subfigure}

\caption{The differential cross sections for $\mu^+\mu^-$ pairs with respect to the invariant mass $M_{\mu\mu}$ in bins of (a) $|y_{\mu\mu}|<0.8$ (b) $0.8<|y_{\mu\mu}|<1.6$ (c) $1.6<|y_{\mu\mu}|<2.4$.  Results for the inclusive, 0n0n, 0nXn, and XnXn configurations are compared with ATLAS Data in Pb+Pb collisions at $\sqrt{s_{NN}}=$5.02 TeV \cite{ATLASdimuon}. The statistical uncertainties of the data (vertical bars) and the respective bin sizes (horizontal bars) are shown.}
\label{fig:atlas_mass_rapidity_bins}

\end{figure*}

\begin{figure*}
\begin{subfigure}{0.48\linewidth}
\centering
\includegraphics[width=\linewidth]{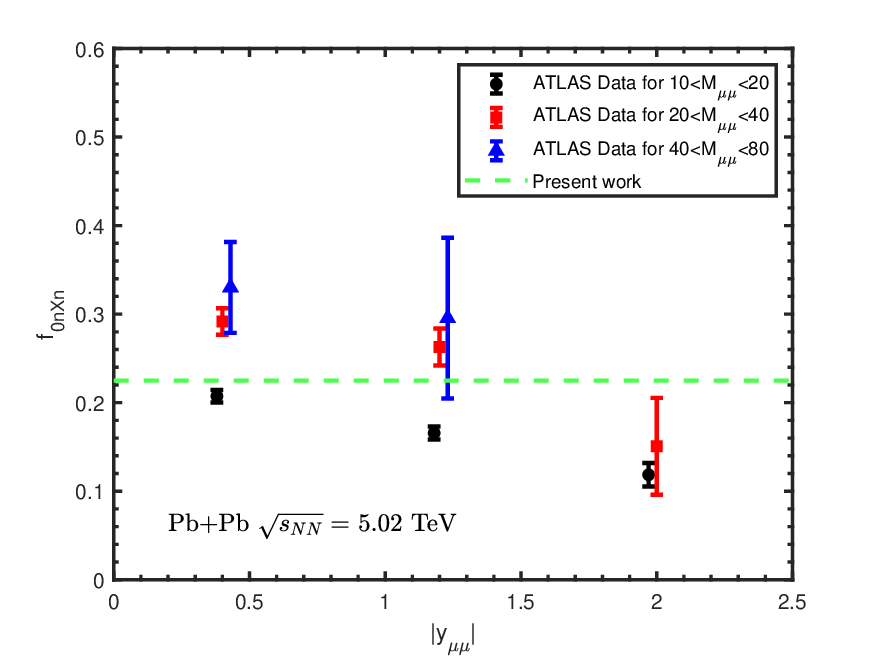}
\caption{}
\label{fig:frac1muon}
\end{subfigure}
\hfill
\begin{subfigure}{0.48\linewidth}
\centering
\includegraphics[width=\linewidth]{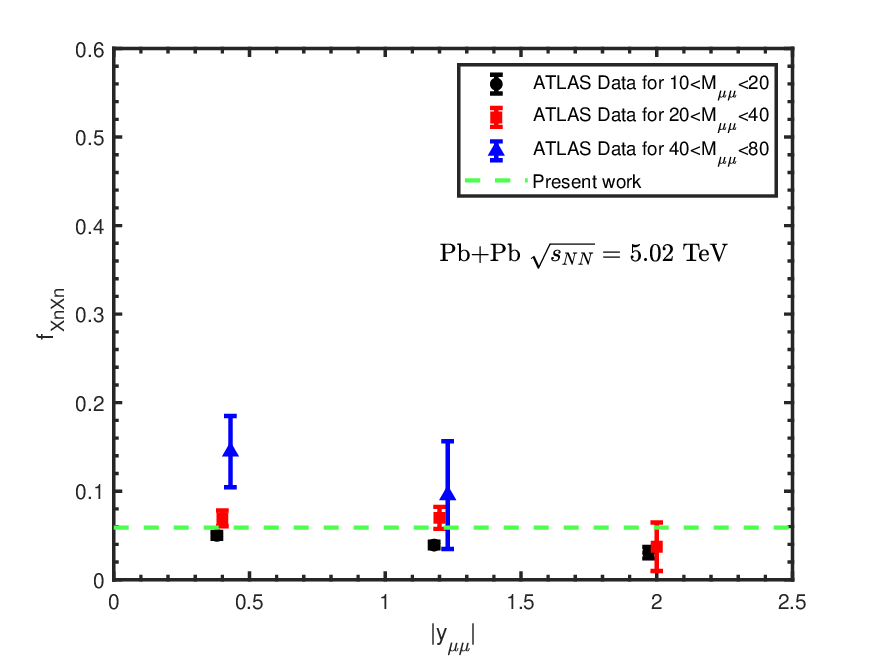}
\caption{}
\label{fig:frac2muonXnXnrapidity}
\end{subfigure}

\begin{subfigure}{0.48\linewidth}
\centering
\includegraphics[width=\linewidth]{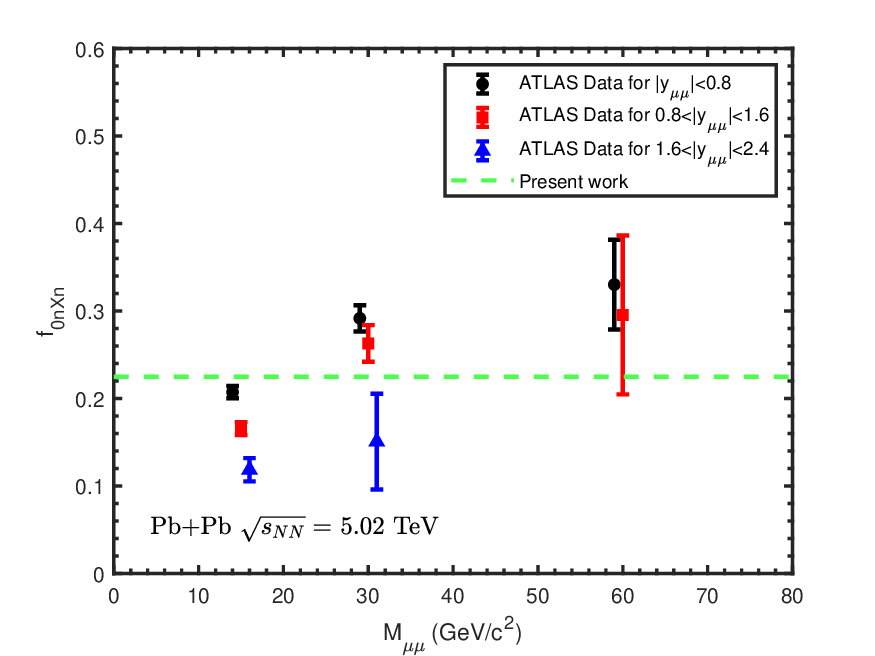}
\caption{}
\label{fig:frac3muon0nXnmass}
\end{subfigure}
\hfill
\begin{subfigure}{0.48\linewidth}
\centering
\includegraphics[width=\linewidth]{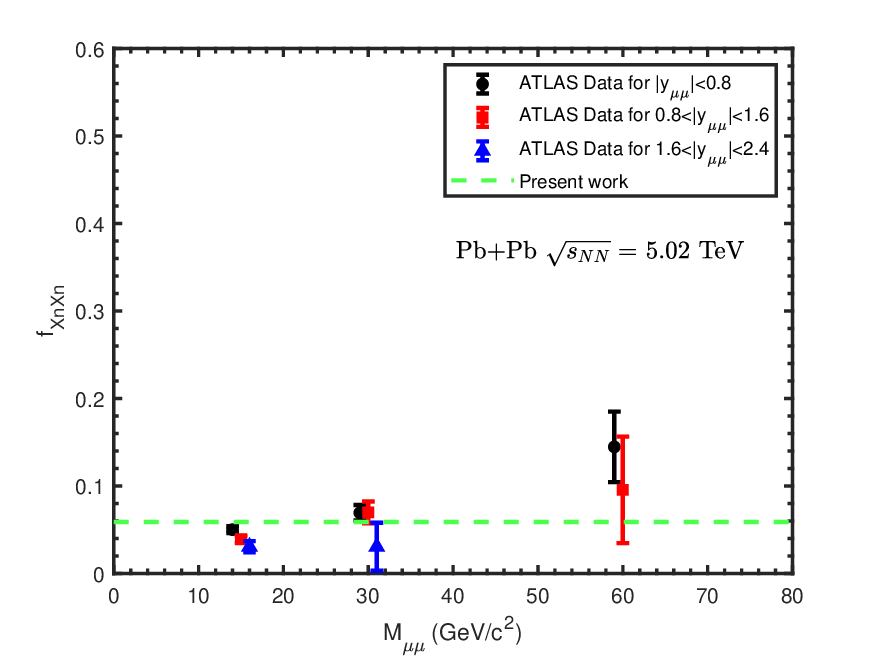}
\caption{}
\label{fig:frac4muonXnXnmass}
\end{subfigure}

\caption{Fractions of the (a) 0nXn and (b) XnXn categories as a function of $M_{\mu\mu}$ in three bins of $|y_{\mu\mu}|$, and of the (c) 0nXn and (d) XnXn categories as a function of $|y_{\mu\mu}|$ in three bins of $M_{\mu\mu}$. The experimental data points and their uncertainties were reproduced from the measurements reported by ATLAS \cite{ATLASdimuon}. Error bars represent statistical uncertainties. The dashed line shows the present work. In (a) and (b), the points corresponding to $|y_{\mu\mu}|<0.8$ and $1.6<|y_{\mu\mu}|<2.4$ are displaced horizontally and in (c) and (d), the points corresponding to $10<M_{\mu\mu}<20$ GeV and $40<M_{\mu\mu}<80$ GeV are displaced horizontally for better visibility.}
\label{fig:fracsAtlasmuon}

\end{figure*}

Experimentally,  ATLAS
Collaboration \cite{ATLASdimuon} has reported a measurement of $\mu^+\mu^-$  production, yielding fiducial cross section of $34.1 \pm 0.3_{\rm stat} \pm
0.7_{\rm sys}$ $\mu$b in ultraperipheral Pb+Pb collisions at 5.02 TeV. Also, the STARLight prediction is 32.1 $\mu$b, while the STARLight+Pythia8 prediction is 30.8 $\mu$b which provides a reference for comparison.

The muon selection is defined within a fiducial phase space where each muon is required to have a transverse momentum $p_T^\mu > 4$ GeV/c and pseudorapidity $|\eta^\mu| < 2.4$. In addition, the dimuon system is restricted within $|y^{\mu\mu}| < 2.4$, with a total transverse momentum $p_T^{\mu\mu} < 2$ GeV/c, and an invariant mass satisfying $M_{\mu\mu} > 10$ GeV/c$^2$
We calculated the cross sections separately for different neutron emission scenarios: 0n0n, 0nXn, XnXn, and inclusive case and present this results in Table \ref{tab:neutron_ATLAS_muon}, where they are compared with the inclusive ATLAS measurement and the corresponding predictions from the STARLight \cite{klein2017starlight} and STARLight+Pythia8 \cite{pythia}. Our inclusive cross section result is $\sigma_{inc}=36.5$ $\mu$b and this is 7.0\% above the measured ATLAS inclusive value, 13.7\% above the STARLight calculation, 18.5\% above the STARLight + Pythia8 calculation.
The decomposition into neutron emission classes that 0n0n channel contributes with a cross section of $26.1\ \mu\mathrm{b}$. The 0nXn and XnXn channels contribute $8.21\ \mu\mathrm{b}$ and $2.15\ \mu\mathrm{b}$, respectively.

In Fig. \ref{fig:rapidity_mass_bins} we show the differential cross sections as a function of the pair rapidity $|y^{\mu\mu}|$ in three invariant mass intervals: $10 < M_{\mu\mu} < 20$ GeV/c$^2$, $20 < M_{\mu\mu} < 40$ GeV/c$^2$, and $40 < M_{\mu\mu} < 80$ GeV/c$^2$ in the top panel. The data points are taken from ATLAS measurements for inclusive channel are also plotted for comparison and their ratio is provided in the bottom panel. Since the differential cross sections are very small, the distributions are presented on a linear rather than a logarithmic scale for a clearer visualization. The assessment can be made from the corresponding ATLAS data to present theory ratio panels, where the ratios remain very close to unity over the entire rapidity range for all invariant mass intervals considered. The deviations from unity are generally small, indicating that the present calculations produce pair rapidity distributions with good accuracy. 

In Fig. \ref{fig:atlas_mass_rapidity_bins} we show the invariant mass distributions in three rapidity intervals: $|y^{\mu\mu}| < 0.8$, $0.8 < |y^{\mu\mu}| < 1.6$, and $1.6 < |y^{\mu\mu}| < 2.4$ in the top panel.
The data points are taken from ATLAS measurements for inclusive channel
are also plotted for comparison and their ratio is provided in the bottom panel. In this case, the vertical axis is presented on a logarithmic scale, making the overall agreement between the calculations and the ATLAS data more apparent across the full range of the distribution. As can be seen from the ratio panels, ATLAS data to present theory ratio remain close to unity up to relatively high invariant masses, indicating a good agreement. At larger invariant masses, however, the deviations become more pronounced. As discussed previously, this behavior is primarily related to the absence of  harder photons $k_\perp$ at small $b$ for the differential cross sections, then the calculation becomes less sensitive at higher masses, resulting in an increasing discrepancy.

In Fig. \ref{fig:fracsAtlasmuon}, the fractions of the 0nXn and XnXn categories are compared with the corresponding experimental measurements of ATLAS as functions of both the invariant mass and the pair rapidity. The present work appear as constant horizontal lines in each panel. This behavior arises from the simplified treatment adopted in the calculation, where the fractions were obtained from the ratios of the corresponding cross sections. Nevertheless, despite this simplified approach, small deviations can be observed in some mass and rapidity intervals, the overall level of agreement remains good.

Additionally, the CMS Collaboration \cite{CMSmuon} has performed a detailed study of neutron-tagged $\gamma\gamma \rightarrow \mu^+\mu^-$ production in Pb+Pb collisions at $\sqrt{s_{\mathrm{NN}}}=5.02$ TeV, in six neutron classes (0n0n through XnXn) and thereby providing experimental information directly relevant to the neutron - tagged dimuon predictions presented in this work.

\subsection{Ditau Production at LHC: Comparison with CMS}
\label{sec:10}

\begin{table*}
\centering
\caption{Neutron tagging of the experimental and calculated results of cross sections for $\tau^+\tau^-$ pair at 5.02 TeV (CMS).}
\label{tab:neutron_classes}
\begin{tabular}{ccc}
\hline\noalign{\smallskip}
Category & Class / Source & Cross section ($\mu$b) \\
\hline

CMS (inclusive) \cite{CMSditau} &  0n0n+0nXn+XnXn
& $4.8 \pm 0.6_{\rm stat} \pm 0.5_{\rm sys}$  \\

CMS extrapolated (inclusive) \cite{shao2025dimuon} &  0n0n+0nXn+XnXn
& 580-850  \\
 
gamma-UPC ChFF (EDFF) (inclusive) \cite{shao2025dimuon} & 0n0n+0nXn+XnXn
& 1060 (860)  \\

\hline
This work (tagged)& 0n0n & 802  \\
This work (tagged) & 0nXn & 195  \\
This work (tagged) & XnXn & 44.7  \\

\hline
This work (inclusive) & 0n0n+0nXn+XnXn & 1043  \\
\hline
\end{tabular}
\end{table*}

\begin{figure*}
\begin{subfigure}{0.48\linewidth}
\centering
\includegraphics[width=\linewidth]{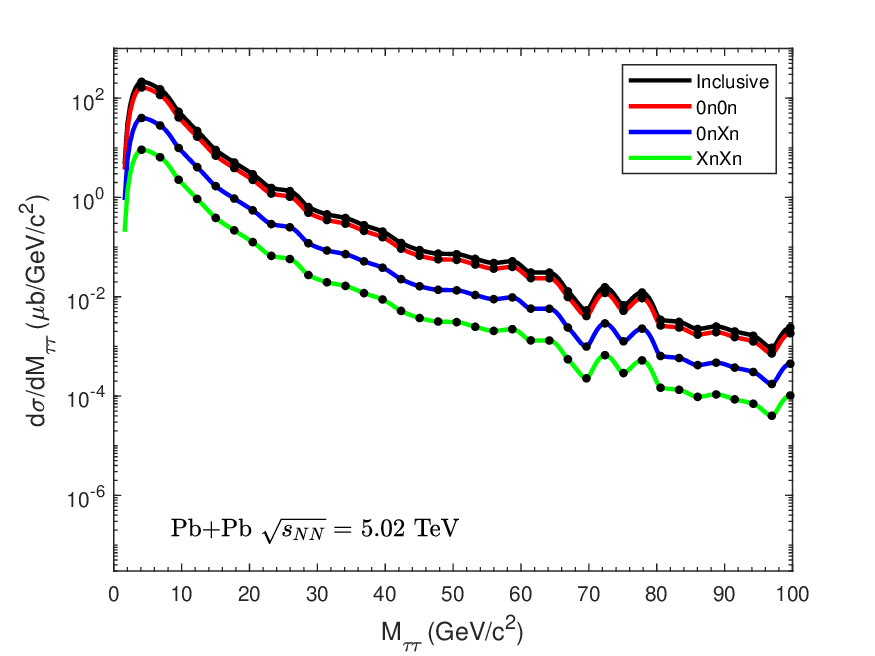}
\caption{}
\label{fig:cms_mass}
\end{subfigure}
\hfill
\begin{subfigure}{0.48\linewidth}
\centering
\includegraphics[width=\linewidth]{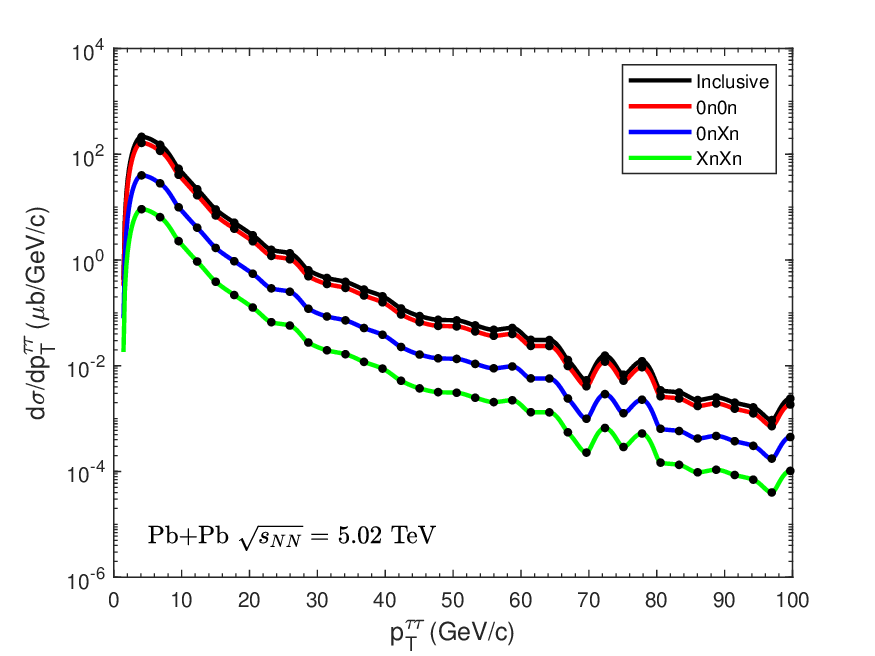}
\caption{}
\label{fig:cms_rapidity}
\end{subfigure}

\caption{The differential cross sections for $\tau^+\tau^-$ pairs with respect to (a) the invariant mass $M_{\tau\tau}$ (b) the pair momentum $p_{T,\tau\tau}$. Results for the inclusive, 0n0n, 0nXn, and XnXn configurations in ultraperipheral Pb+Pb collisions at $\sqrt{s_{NN}}=$5.02 TeV. The statistical uncertainties of the data (vertical bars) are shown.}
\label{fig:cms_distributions}

\end{figure*}

Experimentally, the CMS Collaboration \cite{CMSditau} has reported a measurement of $\tau^+\tau^-$ production in Pb–Pb UPCs at $\sqrt{s_{NN}}=5.02$ TeV in a restricted fiducial phase space, yielding a cross section of $4.8 \pm 0.6_{\rm stat} \pm 0.5_{\rm sys}\ \mu$b. However, due to the leptonic and hadronic decay products of $\tau$ leptons and their detector-level acceptance criteria, this fiducial result cannot be directly compared with theoretical predictions. To enable such a comparison, the measurement is extrapolated to the full phase space. The extrapolation of the CMS fiducial measurement to the full phase space is given in Shao and d'Enterria's work in Ref. \cite{shao2025dimuon} from following Ref. \cite{beresford2020new}, the CMS fiducial result is transformed an inclusive cross section of  580 $\mu\mathrm{b}$, while following Ref. \cite{dyndal2020anomalous} leads to a larger value of 850 $\mu\mathrm{b}$.

On the theoretical side, Shao and d'Enterria calculated the ditau cross sections using Monte Carlo generators such as $\text{MadGraph5\_aMC@NLO}$ combined with dedicated implementations of ultraperipheral photon fluxes. Two commonly used flux models are based on the nuclear charge form factor (ChFF) and the elastic dipole form factor (EDFF), which differ in their treatment of the nuclear charge distribution and consequently in the high energy behavior of the photon spectrum. These calculations yield leading order cross sections of $\sigma_{\mathrm{LO}} = 1060\ \mu\mathrm{b}$ (ChFF) and $\sigma_{\mathrm{LO}} = 860\ \mu\mathrm{b}$ (EDFF), with next-to-leading order QED corrections increasing these values to $\sigma_{\mathrm{NLO}} = 1070\ \mu\mathrm{b}$ and $\sigma_{\mathrm{NLO}} = 870\ \mu\mathrm{b}$, respectively. The corresponding $K$ factor $\frac{\sigma_{NLO}}{\sigma_{LO}}$ remains close to unity ($K = 1.01$), indicating that inclusive $\tau$ pair production is only weakly affected by higher order QED effects.

In addition to these existing predictions, we obtained the cross sections by decomposing into neutron emission classes. Our results summarized in Table \ref{tab:neutron_classes}. The dominant contribution arises from the $0n0n$ channel, with cross section of $802\ \mu$b. The electromagnetic excitation channels, 0nXn and XnXn, contribute $195\ \mu$b and $44.7\ \mu$b, respectively. The inclusive sum over neutron classes, (0n0n + 0nXn + XnXn), yields $1043\ \mu$b, higher than the extrapolated CMS values but in very close agreement with the Shao and d'Enterria's lowest order ChFF result. Further validate the theoretical description, the differential distributions are shown in Fig.  \ref{fig:cms_distributions} with neutron emission. In Fig.\ref{fig:cms_mass}, we show the invariant mass distribution $d\sigma/dM_{\tau\tau}$ and in Fig. \ref{fig:cms_rapidity}, we show the transverse momentum distribution of the pair $d\sigma/dp_T^{\tau\tau}$, without kinematic restriction for inclusive and tagged configurations.

Additionally, $\gamma\gamma \rightarrow \tau^+\tau^-$ production with a no-forward neutron (effectively 0n-tagged) selection, was observed in the ATLAS Collaboration \cite{ATLAStau} and may serve as a useful point of reference for the 0n0n predictions presented in this work.

\section{Conclusion}

We have presented lowest order QED calculations of $\gamma\gamma \rightarrow \ell^+\ell^-$ ($\ell=e,\mu,\tau$) in ultraperipheral
Au+Au collisions at 200 GeV and Pb+Pb collisions at 2.76 and 5.02 TeV. The impact parameter dependent
pair production probability is obtained from a Monte Carlo evaluation of the S-matrix
elements, with the b-dependence recovered by fitting a single exponential to function $F(q)$ and Fourier-Bessel transforming the fit. This is combined with the photonuclear cross section contributions from several excitation mechanisms, including giant dipole resonance, quasi-deuteron, $\Delta$, and high energy continuum using Poisson statistics to
decompose the cross sections into neutron classes (0n0n, 0nXn, XnXn) within fiducial regions matching STAR, ALICE, ATLAS, and CMS.
Furthermore, the probability of no hadronic interaction has been evaluated within the approach of Glauber. 

While neutron tagging has been studied theoretically for electron-positron pair production, its application to heavier lepton pair production remains fewer in the literature. Therefore, one of the primary motivations and key contributions of this study is to provide a simplified theoretical analysis of neutron tagged muon and tau production. 

We have not include higher order QED corrections. Although a detailed photonuclear excitation model has been used to describe electromagnetic dissociation, neutron emission has been treated within a simplified Poisson framework assuming one neutron per excitation. However the results demonstrate that this methods enable a satisfactory description of lepton pair production in UPCs.  It is also important to emphasize that, next-to-leading order QED corrections such as final state radiation, have only a minor impact on the total cross section,  and  particularly relevant for precision studies aiming to extract electromagnetic properties.


\bibliographystyle{spphys}       
\bibliography{references}   

%
%

\end{document}